\documentclass[aip,jcp,reprint ,superscriptaddress, citeautoscript]{revtex4-1}
\pdfoutput=1

\usepackage{amsmath,amsfonts,amssymb,bm,graphicx}
\usepackage{xcolor}

\usepackage[pdftex,bookmarks, pdffitwindow=false, pdfstartview=FitH, pdfdisplaydoctitle, colorlinks, plainpages=false,pdfpagelabels, hypertexnames, citecolor={blue},linkcolor={red}, urlcolor={violet!50!black}, pdflang={en}, hyperfootnotes=false, breaklinks]{hyperref}
\makeatletter
\def\@bibdataout@aip{
 \immediate\write\@bibdataout{
 @CONTROL{
   aip41Control, author="08",editor="1",pages="0",title="0",year="1"
 }}
 \if@filesw
  \immediate\write\@auxout{\string\citation{aip41Control}}
 \fi
}
\makeatother

\begin{document}

\newcommand{\LL}{\mathbb{L}}
\newcommand{\HH}{\mathbb{H}}
\newcommand{\TT}{\mathcal{T}}
\newcommand{\ee}{\mathrm{e}}

\newcommand{\jj}{A}

\newcommand{\ddiv}{\operatorname{div}}

\newcommand{\rhobar}{\overline{\rho}}

\newcommand{\beq}{\begin{equation}}
\newcommand{\eeq}{\end{equation}}

\newcommand{\rlj}[1]{{#1}}
\newcommand{\pbr}[1]{{\color{orange}#1}}

\newcommand{\eqnRef}[1]{Eq.~\eqref{#1}}

\newcommand{\eps}{\epsilon}

\newcommand{\RR}{\bm{R}}
\newcommand{\rr}{\bm{r}}
\newcommand{\CC}{\mathcal{C}}
\newcommand{\FF}{\mathcal{F}}
\newcommand{\NNf}{\mathcal{N}_{\rm f}}
\newcommand{\NNc}{\mathcal{N}_{\rm c}}

\newcommand{\sigB}{\sigma_{\rm B}}
\newcommand{\sigS}{\sigma_{\rm S}}

\newcommand{\muB}{\mu_{\rm B}}
\newcommand{\muS}{\mu_{\rm S}}
\newcommand{\etaS}{\eta_{\rm S}^{\rm r}}

\newcommand{\dd}{\mathrm{d}}

\newcommand{\HK}[1]{{\color{red} #1}}

\title{Correction of coarse-graining errors by a two-level method: application to the Asakura-Oosawa model}

\author{Hideki Kobayashi}
\affiliation{Department of Chemistry, University of Cambridge, Lensfield Road, Cambridge CB2 1EW, United Kingdom}
\author{Paul B. Rohrbach}
\affiliation{Department of Applied Mathematics and Theoretical Physics, University of Cambridge, Wilberforce Road, Cambridge CB3 0WA, United Kingdom}
\author{Robert Scheichl}
\affiliation{Institute for Applied Mathematics, Heidelberg University, Im Neuenheimer Feld 205, 69120 Heidelberg, Germany}
\author{Nigel B. Wilding}
\affiliation{H.H. Wills Physics Laboratory, University of Bristol, Royal Fort, Bristol BS8 1TL, United Kingdom}
\author{Robert L. Jack}
\affiliation{Department of Chemistry, University of Cambridge, Lensfield Road, Cambridge CB2 1EW, United Kingdom}
\affiliation{Department of Applied Mathematics and Theoretical Physics, University of Cambridge, Wilberforce Road, Cambridge CB3 0WA, United Kingdom}

\begin{abstract} 
We present a method that exploits self-consistent simulation of coarse-grained and fine-grained models, in order to analyse properties of physical systems.
The method uses the coarse-grained model to obtain a first estimate of the quantity of interest, before computing a correction by analysing properties of the fine system.
We illustrate the method by applying it to the Asakura-Oosawa (AO) model of colloid-polymer mixtures.  We show that the liquid-vapour critical point in that system is affected by three-body interactions which are neglected in the corresponding coarse-grained model.  We analyse the size of this effect and the nature of the three-body interactions.  We also analyse the accuracy of the method, as a function of the associated computational effort.
\end{abstract}

\maketitle

\section{Introduction}

Coarse-grained models are useful throughout many areas of physics and chemistry~\cite{Likos2001,Pak2018}.  Depending on their context, they are used in different ways.  In some cases, the idea is to develop a simple model that captures important qualitative features of some physical system~\cite{Hagan2006,Haxton2012,Saric2014}.  Alternatively, one may develop a coarse-grained (CG) model that captures quantitatively the behaviour of a more detailed (fine-grained, FG) model~\cite{Noid2008,Praprotnik2007,Ouldridge2011,Mladek2013,Pak2018}.  This second type of application is one of the foundations of multi-scale modelling approaches~\cite{Karplus2014,Warshel2014}.

Enormous theoretical and computational effort has been invested in the development of accurate CG models.  In the simplest case, one starts from a FG model with a classical Hamiltonian, and one aims to characterise its potential energy surface (PES) as a function of some simpler collective variables~\cite{Noid2008,John2017}.  Alternatively, the FG model might be a quantum mechanical system, and one develops a CG model that aims to fit the Born-Oppenheimer PES for the nuclear co-ordinates~\cite{Karplus2014,Warshel2014}.  A number of recent works have used machine-learning methods~\cite{Behler2007,Bartok2010,Cheng2019} to derive approximate PESs in this case.  

Independent of the specific application, the most common approach in such cases is to derive a CG model that is as accurate as possible, and then to estimate observable properties of the original (FG) model by computing the corresponding quantities in the CG model.  If the CG model is an accurate representation of the fine one then this approach works well.  However, it is very rare for the CG model to be perfect, which means that this approach almost always results in some systematic errors, when estimating properties of the fine model.  No information about these errors is available from analysis of the CG model alone.

In this work, we discuss a method for eliminating these coarse-graining errors.  
Specifically, we make quantitative estimates of observable properties of the FG model, which we decompose as the value of the corresponding quantity for the CG model, and an explicit correction term.  We show how this correction can be estimated, in an efficient way.  Our method is inspired by recent mathematical studies, particularly the multi-level approach~\cite{heinrich1998monte,Giles2008,Anderson2012}, which has been developed recently for applications in Bayesian statistics and uncertainty quantification~\cite{Hoang2013,Dodwell2015,beskos2017multilevel}.  (We note that the term ``multi-level'' has a specific mathematical meaning and should not be confused with the ``multi-scale'' terminology used in physics and chemistry~\cite{Karplus2014,Warshel2014}.)  Our method corresponds to a multi-level implementation with only two levels, so we call it a two-level (TL) method.  It shares many features with importance sampling methods, biased sampling techniques~\cite{frenkel-smit,Bruce2003}, and free-energy perturbation methods~\cite{Zwanzig1954} (see for example Ref.~\onlinecite{Cheng2019} for a recent application of this method, in order to correct for coarse-graining errors).  
 Our implementation of the TL method is informed by recent mathematical results and has been chosen to minimise the systematic errors (bias) in our estimates of properties of the FG system.  As we discuss below, we minimise the bias by using a method for free-energy estimation~\cite{Hummer2001} that is based on the fluctuation theorems of Jarzynski~\cite{Jarzynski1997} and Crooks~\cite{Crooks2000}.  
This free-energy estimation method is equivalent to the mathematical technique of annealed importance sampling~\cite{Neal2001}. 

The method that we present is general.  As a specific application, we use it to analyse the Asakura-Oosawa (AO) model~\cite{Asakura1954,Binder2014} of colloid-polymer mixtures~\cite{Poon2002}.  In this model, colloidal particles and polymers are described as spheres of different sizes, with pairwise interactions.  The ratio of particle sizes is $q \leq 1$.  In the corresponding CG model, only the colloidal particles are included.  If $q$ is sufficiently small then a CG model with pairwise interactions provides an exact description of the colloidal degrees of freedom~\cite{Dijkstra1999-jpcm}.  For larger $q$, only approximate CG models are available, because of effective interactions among the large particles that cannot be described by a pairwise additive potential~\cite{Dijkstra1999-jpcm,Loverso2006,Ashton2014-jcp}.  In this case -- as in many other CG models which restrict to pairwise interactions~\cite{Likos2001,Ashton2011depletion} -- corrections to the CG model are required~\cite{Ashton2014-jcp} in order to accurately estimate properties of the full (fine-grained) mixture.  

\rlj{For the specific case of the AO model, a method for calculation of these corrections was previously exploited by Dijkstra and co-workers~\cite{Dijkstra2002,Dijkstra2006}.   They used a geometrical construction to analyse many-body interaction terms on the fly, during a CG simulation.  Our method provides an alternative route to computation of these corrections: it provides an accurate characterisation of the critical point of the AO mixture.   As expected~\cite{Dijkstra2002,Dijkstra2006,Loverso2006}, we} find that many-body effective interactions for the colloids tend to reduce the tendency of the system to form clusters.  We also analyse the nature and strength of the three-body and many-body effective interactions.  In particular, for $q=\frac25$, we show that a CG model with two- and three-body interactions can capture the critical point to high accuracy, where a two-body CG model fails.  

The structure of the paper is as follows.  Sec.~\ref{sec:method-general} describes the general method, while Sec.~\ref{sec:ao} describes the AO model and details how the method is applied in this case.  In Sec.~\ref{sec:results} we present numerical results.  Then, Sec.~\ref{sec:convergence} analyses the accuracy of the method, and its statistical uncertainties.  Finally, Sec.~\ref{sec:conc} summarises the conclusions and outlook.  Additional technical detail is provided in the appendices.

\section{The  Two-level Method}
\label{sec:method-general}

\subsection{Outline}

This  Section presents our method in general terms, using a schematic notation.  The specific application that we consider is a mixture of large and small particles, for which full definitions are given in Sec.~\ref{sec:ao}.

\begin{figure}
\includegraphics[width=80mm]{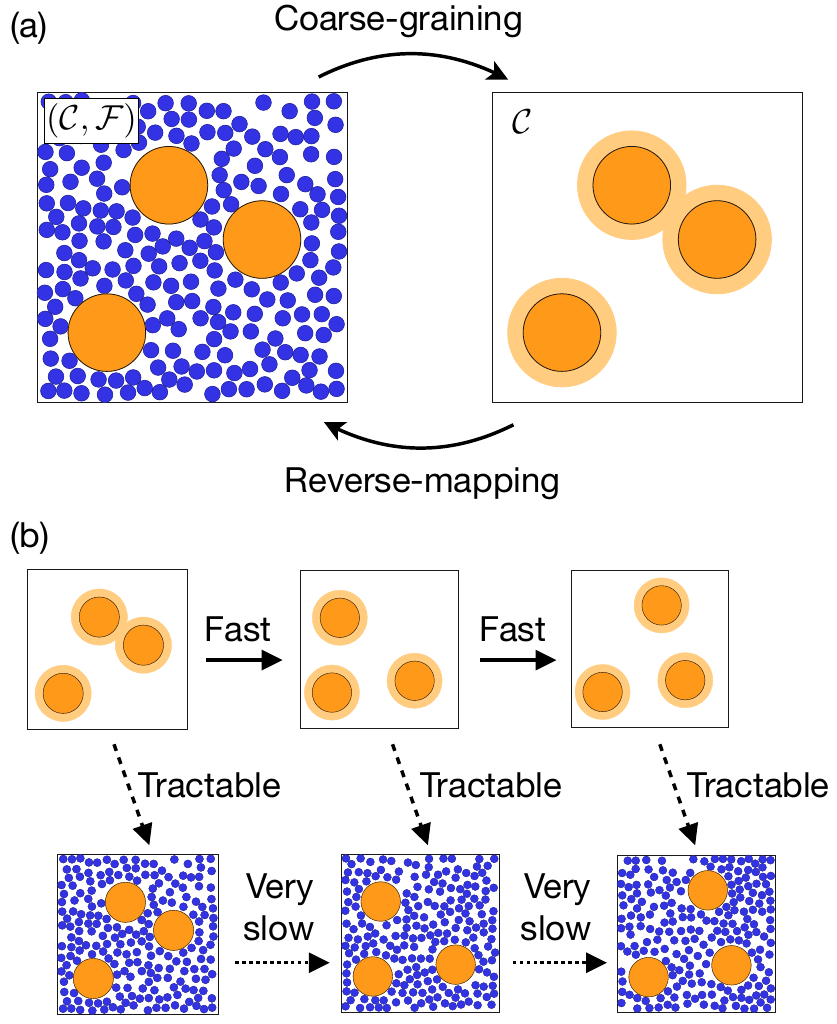}
\caption{(a) Illustration of a FG mixture of large and small particles, and a corresponding CG model that includes only the large particles, interacting by an effective potential. (b)~TL~method.   Direct MC simulation of the FG model is very slow, so the TL method bypasses this step.  It uses (fast) simulation of the CG model to generate independent configurations of the large particles.  The small particles are then repopulated via a tractable step (similar to a reverse-mapping) that also yields an estimate for their free energy.  This enables accurate computation of averages in the FG system.}
\label{fig:schematic}
\end{figure}

Consider a system with many degrees of freedom, and suppose that we are given a coarse-grained model which describes the same physical system in terms of a smaller set of (coarse) variables.
In our application, the full system consists of a mixture of large and small particles, in the grand canonical ensemble.  The coarse degrees of freedom are  the number of large particles $N$, and their positions $(\RR_1,\RR_2,\dots,\RR_N)$.  The other degrees of freedom are  the number of small particles $n$, and their positions $(\rr_1,\rr_2,\dots,\rr_n)$.  In our schematic notation, we denote a configuration of the coarse system as $\CC=(N,\RR_1,\RR_2,\dots,\RR_N)$ and a configuration of the full system as $(\CC,\FF)$ with $\FF=(n,\rr_1,\rr_2,\dots,\rr_n)$.  This full system is the FG system, whose degrees of freedom comprise both $\CC$ and $\FF$.  See Fig.~\ref{fig:schematic}(a).

Given some configuration of this FG system, the energy is $E(\CC,\FF)$ and the associated Boltzmann distribution is 
\beq
p_{\rm f}(\CC,\FF) = \frac{1}{Z}  \ee^{-\beta E(\CC,\FF)}
\label{equ:pf}
\eeq
where $\beta=1/(k_{\rm B}T)$ is the inverse temperature (scaled by Boltzmann's constant $k_{\rm B}$), and $Z$ is the partition function.  

The aim of the method is to compute averages with respect to $p_{\rm f}$.  We restrict to averages of observable quantities that depend only on the coarse degrees of freedom.  The average of such a quantity $A$ is denoted by 
\beq
\langle A \rangle_{\rm f} = \int p_{\rm f}(\CC,\FF) A(\CC) \,\dd\CC\dd\FF \; .
\label{equ:Af}
\eeq
The definition of the integration measure $\dd\CC\dd\FF$ depends on the system of interest.
The marginal probability density for the coarse variables is
\beq
p_{\rm m}(\CC) = \int p_{\rm f}(\CC,\FF) \, \mathrm{d}\FF \; .
\label{equ:pm}
\eeq
Hence $\langle A \rangle_{\rm f}  = \int p_{\rm m}(\CC) A(\CC) \dd \CC$. 

A configuration $\CC$ of the coarse-grained system has energy $E_{\rm c}(\CC)$.  The Boltzmann distribution for the coarse system is
\beq
p_{\rm c}(\CC) = \frac{1}{Z_{\rm c}}  \ee^{-\beta E_{\rm c}(\CC)}
\label{equ:pc}
\eeq
and the associated equilibrium average is
\beq
\langle A \rangle_{\rm c} = \int p_{\rm c}(\CC) A(\CC) \,\dd\CC  \; .
\label{equ:Ac}
\eeq
For a given fine system, the identification of suitable coarse degrees of freedom $\CC$ and the inference of the energy function $E_{\rm c}(\CC)$ are both part of the coarse-graining procedure.  A related procedure is the generation of representative configurations of the full (FG) system, starting from coarse configurations $\CC$; this is known as reverse-mapping, see for example Ref.~\onlinecite{Spyriouni2007}.   See Fig.~\ref{fig:schematic}(a).

A great deal of effort~\cite{Likos2001,Behler2007,Bartok2010,Ouldridge2011,Mladek2013,Pak2018} has gone into developing energy functions $E_{\rm c}$ such that coarse averages such as $\langle A \rangle_{\rm c}$ provide good estimates of the corresponding fine average $\langle A\rangle_{\rm f}$.  This obviously requires that $p_{\rm c} \approx p_{\rm m}$.
Our method addresses the following question:  Suppose that $E$ and $E_{\rm c}$ are given, with $p_{\rm c}$ close (but not equal) to $p_{\rm m}$.  How should one obtain an accurate estimate of $\langle A\rangle_{\rm f}$, at reasonable computational cost?  

The TL method can achieve this, as shown schematically in Fig.~\ref{fig:schematic}(b).   It combines computations on the CG model with a reverse-mapping procedure that is used to assess the accuracy of the this model.  We first estimate $\langle A\rangle_{\rm c}$ by classical Monte Carlo (MC) sampling, which is assumed to have a relatively low computational cost.  (In mathematics, this would be called a Markov chain Monte Carlo (MCMC) method.)  This method is used to generate  $\NNc$ representative samples from (\ref{equ:pc}) which should be identically distributed but need not be independent.  These are denoted by $(\CC_1,\CC_2,\dots,\CC_{\NNc})$.  Define the estimator
\beq
\hat{A}_{\rm c} = \frac{1}{\NNc} \sum_{i=1}^{\NNc} A(\CC_i) \; .
\label{equ:hat-Ac}
\eeq
This is a consistent estimator of $\langle A\rangle_{\rm c}$ but it is not an accurate estimate of the quantity of interest, which is $\langle A\rangle_{\rm f}$.  To quantify the difference between these two averages
we make a computational estimate of  
\begin{align}
\langle A \rangle_{\rm f} - \langle A \rangle_{\rm c} & = \int \left[ \frac{ p_{\rm m}(\CC)}{ p_{\rm c}(\CC)} - 1 \right] A(\CC) p_{\rm c}(\CC) \, \dd \CC
\nonumber \\
& = \left\langle [w(\CC) - 1] A(\CC) \right\rangle_{\rm c}
\label{equ:fg-cg}
\end{align}
where $w(\CC) = p_{\rm m}(\CC)/p_{\rm c}(\CC)$ is a reweighting factor.   This is achieved by drawing $\NNf$ samples from those that were used in  (\ref{equ:pf}).  With slight abuse of notation we denote these by $(\CC_1,\CC_2,\dots,\CC_{\NNf})$.  Then one estimates (\ref{equ:fg-cg}) as 
\beq
\hat{\Delta} = \frac{1}{\NNf} \sum_{i=1}^{\NNf} [\hat{w}(\CC_i)-1] A(\CC_i)
\label{equ:hat-Delta}
\eeq
where $\hat{w}(\CC_i)$ is an estimate of the ratio $p_{\rm m}(\CC_i)/p_{\rm c}(\CC_i)$ that is obtained by a free energy computation using Jarzynksi's equality (see Section~\ref{sec:jarz-ao}, this step has significant computational cost but is tractable in the systems of interest).
Finally, one has an estimator for the fine average: 
\beq
\hat{A} = \hat{A}_{\rm c} + \hat{\Delta} \; .
\label{equ:Acf}
\eeq
We show below that as $\NNc,\NNf\to\infty$ then 
\beq 
\hat{A} \to \langle A \rangle_{\rm f} \; .
\label{equ:A-conv}
\eeq

\subsection{General properties}

From (\ref{equ:pm}), one identifies $(-1/\beta) \log p_{\rm m}(\CC)$ as the free energy of the $\FF$ variables, for a given coarse configuration $\CC$.  
Hence $E_{\rm c}(\CC)=(-1/\beta) \log [Z_{\rm c}p_{\rm c}(\CC)]$ is an estimate of this free energy, and the correction  (\ref{equ:Acf}) is analogous to the free-energy perturbation formula of Zwanzig~\cite{Zwanzig1954}.   In general, this perturbation formula suffers from large errors, except when the correction is small.  Our TL method provides a method for computing the weights $\hat{w}$ so as to obtain accurate estimates for $\langle A\rangle_{\rm f}$.

It is natural to compare the TL method with simple MC algorithms that operate directly on the fine system.  
In cases where coarse-graining is useful, one should expect a separation of length scales between coarse and fine degrees of freedom.  Often, this means that direct MC calculations on the fine system are frustrated by the disparate length scales.  For example, it is likely that MC moves should involve co-ordinates that change only on the fine scale, but exploration of configuration space requires changes over the (much larger) coarse scale.  The result is that it takes many such moves to relax (or decorrelate) the configurations of the system.   In mathematical parlance, the MCMC chain for the FG system is expected to mix very slowly.  

The TL method avoids this problem: the central idea is illustrated in Fig.~\ref{fig:schematic}(b).  Note that the estimator ${\hat A}_{\rm c}$ only involves coarse degrees of freedom, so it does not suffer from the separation of length scales.  Computation of ${\hat \Delta}$ requires MC computations to be performed on the fine degrees of freedom but it does not require that the coarse degrees of freedom are relaxed during this process.  Hence, this computation does not suffer from the (very slow) mixing in the FG system.  In cases where the CG model is accurate, this can lead to a strong improvement in performance.

Given a FG system, the TL method can be applied using any CG model.  The results are (in principle) correct in all cases.  This feature -- which is standard within methods based on histogram reweighting, importance sampling, and free-energy perturbations~\cite{Bruce2003,Zwanzig1954} -- allows enormous flexibility in the choice of CG model.  However, if the CG model is a poor representation of the fine one, the method is not efficient from a numerical point of view.  In this case, the weights $\hat{w}_i$ acquire large fluctuations, which leads to large statistical errors on $\hat\Delta$ (that is, large error bars).  This is an in-built diagnostic of the quality of the CG model, within the method.  The computed weights can also be used to refine CG models, to improve their accuracy.

\section{Asakura-Oosawa model system}
\label{sec:ao}

\subsection{Fine-grained system}

\begin{figure}
\includegraphics[width=80mm]{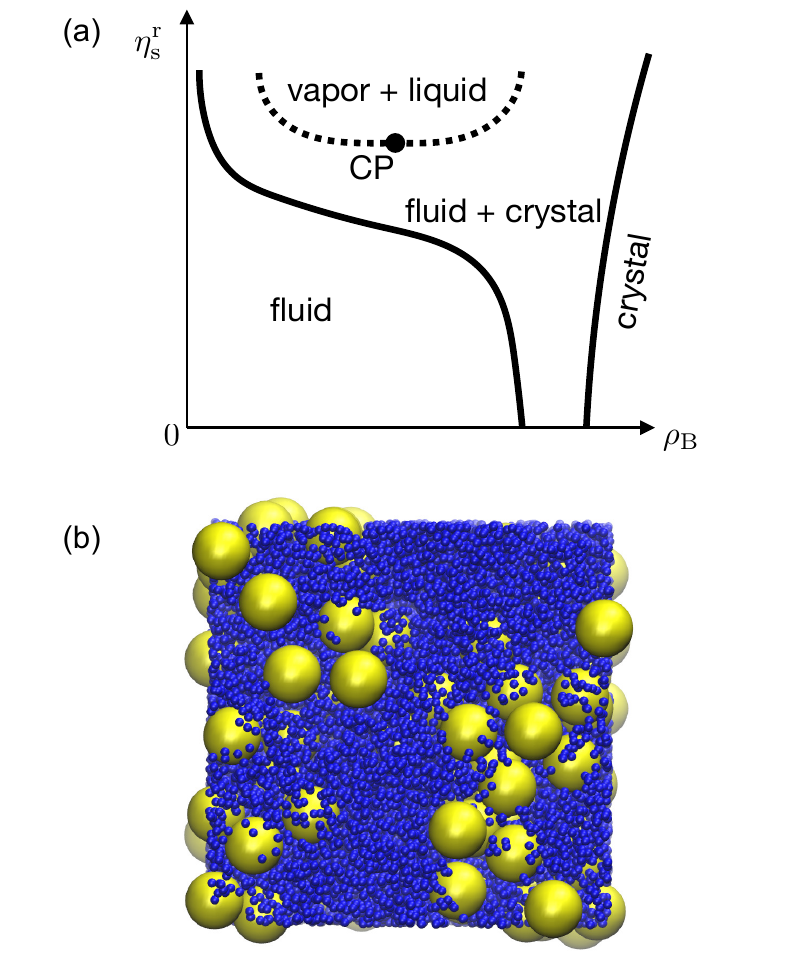}
\caption{(a) Sketch of the phase diagram for the AO system~\cite{Dijkstra1999-jpcm,Binder2014}, as a function of the number density of large particles $\rho_{\rm B}$ and the reservoir volume fraction of small particles $\etaS$.  For $q\lesssim0.4$, the liquid-vapour binodal occurs inside the fluid-crystal coexistence regime and is therefore metastable~\cite{Dijkstra1999-jpcm}.  This work focusses on the (metastable) critical point marked by CP, which signals the onset of phase coexistence of the colloidal liquid and colloidal vapor phases.
(b) Snapshot of the fine system at its critical point.  The size ratio is $q=(2/13)\approx 0.1538$ which is small enough that the coarse model is exact.  The box size is $L=6\sigB$ and $\etaS=0.3198$, the number of large particles in this configuration  is $N=108$.
 Direct simulation of this system is challenging due to the size disparity between large and small particles.}
\label{fig:snapshot-phase}
\end{figure}

As an application of the TL method, we consider a model of a colloid-polymer mixture, as studied by Asakura and Oosawa~\cite{Asakura1954} and by Vrij~\cite{Vrij1976}
It consists of large and small spherical particles in three spatial dimensions, with diameters $\sigB$ and $\sigS$ respectively.   
The behaviour of this model is reviewed in Ref.~\onlinecite{Binder2014}, see also Fig.~\ref{fig:snapshot-phase}.
\rlj{Its phase behaviour  has been studied extensively, both computationally~\cite{Dijkstra1999-jpcm,Dijkstra2006} and theoretically~\cite{Gast1983,Lekkerkerker1992,Schmidt2002,Brader2003,Oversteegen2005}.  }
The large particles are hard spheres that represent the colloid.  The small particles are schematic representations of (ideal) polymer chains and their size is given by the polymer radius of gyration.  These spheres are free to overlap with each other, because the excluded volume of a polymer is very small in comparison with its radius of gyration.  In colloid-polymer mixtures, the effect of the polymers is to promote clustering of the colloidal particles, which leads eventually to phase separation and/or crystallisation~\cite{Poon2002}.  The AO model captures both these features~\cite{Binder2014}.  We focus here on fluid phases.

The particles occupy a periodic box in three dimensions, of size $L\times L\times L$ . The large particles are hard: the potential energy $U_{\rm f}(\CC)=\infty$ if two large particles overlap or if a large particle overlaps with a small one; otherwise $U_{\rm f}(\CC)=0$.  This means in particular that small particles are free to overlap with each other.  The chemical potentials of the large and small particles are $\muB,\muS$.  
The Boltzmann distribution for this system is similar to~(\ref{equ:pf}): 
\beq
p_{\rm f}(\CC,\FF) = \frac{1}{\Xi_{\rm f}}\ee^{\beta\muB N + \beta\muS n - \beta U_{\rm f}(\CC,\FF)}
\label{equ:pf-ao}
\eeq
where $\Xi_{\rm f}$ is the grand partition function.  Computation of averages like (\ref{equ:Af}) requires a definition of the integration measure $\mathrm{d}\CC \mathrm{d}\FF$: this is chosen such that the average in (\ref{equ:Af}) is equal to an average over the grand canonical ensemble.  It includes sums over $N,n$ and also a combinatorial factor of $N!n!$ that accounts for particle indistinguishability.  Details are given in Appendix~\ref{app:gce}.

In place of the parameter $\beta\muS$, it is convenient to work with the volume fraction of the small particles $\etaS$ in a (nominal) reservoir at chemical potential $\muS$, that is $\etaS = \pi\ee^{\beta \muS}/6$.  The size ratio between large and small particles is
\beq
q = (\sigS/\sigB) \; .
\eeq

Since the small particles do not interact with one another, the integral in (\ref{equ:pm}) can be computed, by summing over the number of small particles and integrating their positions.  \rlj{Following Ref.~\onlinecite{Dijkstra1999-jpcm}, we show} in Appendix~\ref{app:gce} that for configurations $\CC$ without overlapping particles the result is
\beq
p_{\rm m}(\CC) = \frac{1}{\Xi_{\rm f}}\exp\left[ \beta\muB N  + \frac{6\etaS}{\pi\sigS^3} {\cal V}_{\rm a}(\CC) \right] \; ,
\label{equ:pm-ao}
\eeq
where ${\cal V}_{\rm a}(\CC)$ is the total volume that is accessible to the (centres of) the small particles.

We summarise the dimensionless model parameters, which are the colloid size ratio $q$, the large-particle chemical potential $\beta\muB$, the volume fraction of the small particles (in the reservoir) $\etaS$, and the size of the simulation box $L/\sigB$.

\subsection{Coarse-grained model}

The coarse system only has large particles.  These still have (repulsive) hard cores but they also interact by a pairwise additive  interaction with the Asakura-Oosawa (AO) pair potential.  That is, for configurations $\CC$ without overlapping particles one has~\cite{Binder2014}
\beq
\beta U_{\rm c}(\CC) =\sum_{i=1}^{N-1} \sum_{j=i+1}^N \beta V_{\rm AO}(|\RR_i - \RR_j|)
\label{equ:Ucoarse}
\eeq
with 
\beq
\beta V_{\rm AO}(r) = 
\frac{ -\etaS }{ q^3 }
\left[  (1+q)^3-
\frac{3r(1+q)^2}{2\sigB}
+ \frac{r^3}{2\sigB^3}
\right] \; .
\eeq
If there are overlapping particles then $U_{\rm c}=\infty$.  The corresponding Boltzmann distribution is
\beq
p_{\rm c}(\CC) = \frac{1}{\Xi_{\rm c}} \ee^{\beta\muB N  - \beta U_{\rm c}(\CC)} \; ,
\label{equ:pc-ao}
\eeq
where $\Xi_{\rm c}$ is the coarse partition function.
Comparing (\ref{equ:pm-ao},\ref{equ:pc-ao}), one sees that the coarse model corresponds to a pairwise additive approximation for the accessible volume ${\cal V}_{\rm a}$.
The model is parameterised by the same dimensionless parameters as the fine system ($q,\beta\mu_{\rm B},\etaS,L/\sigB$).

The behaviour of this system depends on the size ratio $q$.  In particular, for $q\leq (2/\sqrt{3})-1 \approx 0.1547$, a geometrical construction can be used to show that the pairwise-additive approximation for ${\cal V}_{\rm a}$ is exact and 
$p_{\rm c}(\CC) = p_{\rm m}(\CC)$ for all coarse configurations  $\CC$~\cite{Binder2014}.  For larger values of $q$, the effective interaction  in the fine system includes 3-body and higher contributions, which means that
\begin{multline}
\frac{p_{\rm m}(\CC) }{ p_{\rm c}(\CC) } \propto \exp\Big(  \beta N \Delta\mu  - \sum_{ijk} \beta V_3(\RR_i,\RR_j,\RR_k) 
\\ - \sum_{ijkl}  \beta V_4(\RR_i,\RR_j,\RR_k,\RR_l)  
 - \dots \Big)
 \label{equ:pm-pc}
\end{multline}
where the sums run over all the large particles, $\Delta\mu$ is a constant, and $V_3,V_4,\dots$ are many-body interaction potentials. 
\rlj{For the specific case of the AO model, these many-body terms can be computed (numerically) using a geometrical construction, as shown in Refs.~\onlinecite{Dijkstra2002,Dijkstra2006}.
For the purposes of this work} we note that $p_{\rm m} \neq p_{\rm c}$ so computations on the coarse system (\ref{equ:pc-ao}) do not provide accurate estimates of properties of the fine system.  Hence the method of Sec.~\ref{sec:method-general} is applicable. 

The estimators (\ref{equ:hat-Ac},\ref{equ:hat-Delta}) require that we generate representative samples from (\ref{equ:pc-ao}).  This is 
 achieved by grand canonical MC simulations of the coarse system~\cite{frenkel-smit}.

\subsection{Observable quantities}
\label{sec:obs}

The phase diagram for the AO system is sketched in Fig.~\ref{fig:snapshot-phase}(a), as a function of $\etaS$ and the large particle number density $\rho_{\rm B}$, for a representative value of $q$.
For small $\etaS$ there is a fluid phase at low to moderate density, and a crystalline phase at high density.  On increasing $\etaS$ at moderate $\rho_{\rm B}$, the crystalline phase becomes stable, but the fluid is still accessible as a metastable phase.
On further increasing $\etaS$ within the metastable fluid, there is a critical point at $(\etaS)_{\rm c}$, and for $\etaS>(\etaS)_{\rm c}$ one observes coexistence between phases with many large particles (``colloidal liquid'') and few large particles (``colloidal vapour'').  
This work concentrates on the behaviour near the critical point. In this regime the system depends very sensitively on interaction parameters, so it provides a challenging test case for the TL method.  
In our grand-canonical MC simulations of the coarse system, the fluid is sufficiently metastable that we do not observe crystallisation (see also Appendix~\ref{app:sim_details}).

In the following we focus on the critical point.  
As a quantity of interest, we consider the histogram of the number of large particles. That is, we compute the probability to have exactly $N_{\rm b}$ large particles, which is
\beq
P_{\rm b}(N_{\rm b}) = \langle \delta_{N,N_{\rm b}} \rangle_{\rm f} \; .
\eeq
This is of the required form (\ref{equ:Af}).
The distribution $P_{\rm b}$ has a characteristic form at the critical point~\cite{wilding1995critical}.  In the vicinity of this point, it depends strongly on the model parameters.  Hence the accurate determination of $P_{\rm b}$ is a stringent test of our method and allows us to locate the critical point of the fine system.

\rlj{A previous study~\cite{Loverso2006} has} characterised this critical point by direct Monte Carlo computations on the fine system.  However, despite the use of advanced MC methods~\cite{Vink2004}, that work was limited to relatively large $q\geq \frac25$, otherwise the scale separation between the large and small particles makes such computations extremely expensive.  Fig.~\ref{fig:snapshot-phase} shows a typical configuration of the FG system near the critical point, for a small size ratio $q=\frac{2}{13}$.  This picture illustrates that particle insertion MC moves for the large particles would be extremely challenging in this regime.  Here we are able to obtain accurate results for the fine system at $q=\frac14$.  
\rlj{This allows us to estimate effects of three-body interactions, which are not captured by the coarse model, although they could be analysed following Ref.~\onlinecite{Dijkstra2006}. }  Our method is also accurate at $q=\frac{2}{13}\approx 0.15$: this case is useful as a test of the method but the coarse model is already exact at that state point so it does not yield any new physical insight.

\subsection{Jarzynski integration and estimation  of $P_{\rm b}$}
\label{sec:jarz-ao}

\begin{figure*}
\includegraphics[width=180mm]{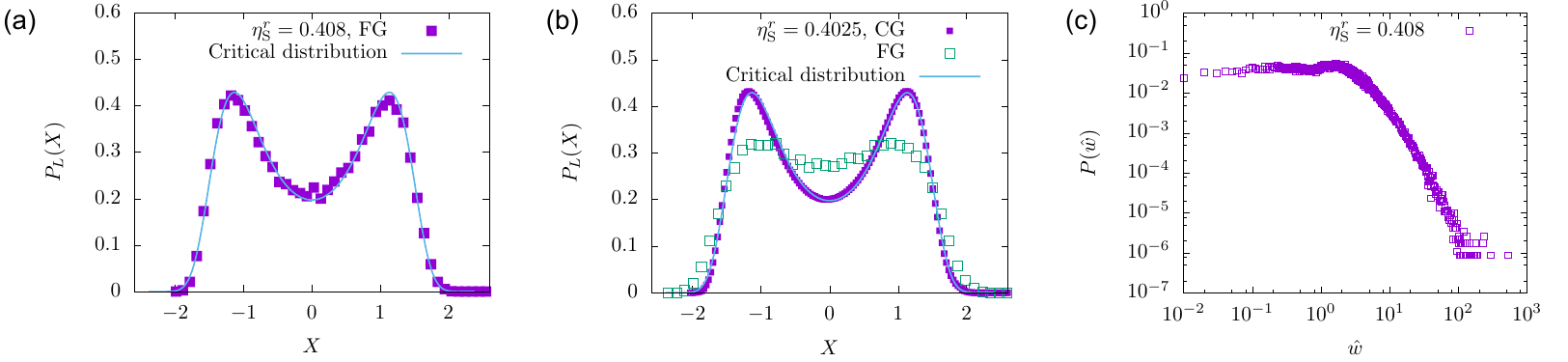}
\caption{Results for $q=(1/4)$ and $L=5\sigB$.
(a)~Histogram of the large-particle density in the fine system (FG) at $\etaS = 0.4080$, compared with the (universal) critical distribution.  This result was obtained using the TL estimator $\hat{A}$ defined in (\ref{equ:Acf}).
(b)~Histogram of the large-particle density in the coarse system (CG) at its critical point ($\etaS\approx 0.4025$), and the corresponding (reweighted) distribution for the fine system (FG) at the same value of $\etaS$.  From the shape of this distribution, one sees that the fine system is in the one-phase regime.
(c)~Histogram of the reweighting factors $\hat{w}_i$ that are used in the estimator (\ref{equ:hat-Delta}) when computing the results in panel (a).}
\label{fig:ratio4}
\end{figure*}

Computation of the weights $\hat w$ in (\ref{equ:hat-Delta}) involves grand canonical MC simulations where the large particles are held fixed while small particles are inserted and removed.  The system is divided into cells of size $\sigS$ so that there are $n_{\rm cell}$ cells in total: this facilitates the computation of whether an attempted small-particle insertion is successful.  We define a Monte Carlo sweep (MCS) as a sequence of $n_{\rm cell}$ attempted MC updates, where each update is chosen to be either a particle insertion or a particle removal, with equal probability.

The weight factors $\hat w(\CC_i)$ are computed in two stages.
We first calculate non-normalised weights $\hat{W}(\CC_i)$ by a Monte Carlo (Jarzynski) integration method~\cite{Hummer2001,Neal2001}. The procedure is described below where we also explain that for any fixed $\CC$  then
\beq
\langle \hat{W}(\CC) \rangle_{\rm J} = \exp\left[\frac{6 \etaS}{\pi\sigS^3} {\cal V}_{\rm a}(\CC) + \beta U_{\rm c}(\CC) \right] \; .
\label{equ:jarz-result}
\eeq 
The average in this equation is an expectation value with respect to the Monte Carlo (Jarzynski) integration (see below).
Then we take
\beq
\hat{w}(\CC_i) = \frac{ \hat{W}(\CC_i) }{ \frac{1}{\NNf} \sum_{j=1}^{\NNf} \hat{W}(\CC_j) } \; .
\label{equ:ratio}
\eeq
Observe from (\ref{equ:pm-ao},\ref{equ:pc-ao}) that (\ref{equ:jarz-result}) is equal to $(\Xi_{\rm f}/\Xi_{\rm c})\cdot(p_{\rm m}(\CC)/p_{\rm c}(\CC))$.
Taking the average of (\ref{equ:ratio}) and comparing with (\ref{equ:pm-ao},\ref{equ:pc-ao}), one has for large $\NNf$ that
\beq
\langle \hat{w}(\CC) \rangle_{\rm J} \approx \frac{p_{\rm m}(\CC)}{p_{\rm c}(\CC)}  \; ,
\label{equ:mean-w}
\eeq
where we used that the denominator of (\ref{equ:ratio}) converges for large $\NNf$ to the expectation of (\ref{equ:jarz-result}) with respect to $p_{\rm c}$, which is $(\Xi_{\rm f}/\Xi_{\rm c})$.  [The approximate equality in (\ref{equ:mean-w}) is accurate up to a small correction associated with the difference between the denominator of (\ref{equ:ratio}) and its average, see Sec.~\ref{sec:convergence}, below.]

We now describe the computation of the weights $\hat{W}$.  Given some $\CC$, we identify the grand partition function for the small particles as
\beq
\Xi_{\rm S}(\muS,\CC) =  \exp\left[ \ee^{\beta\muS}  {\cal V}_{\rm a}(\CC) / \sigS^{3} \right]  = \exp\left[\frac{6 \etaS}{\pi\sigS^3}  {\cal V}_{\rm a}(\CC) \right] \; .
\label{equ:XiS}
\eeq
To compute $\hat{W}(\CC)$, we hold the large particles fixed in position and we perform grand-canonical MC for the small particles.
We first choose some chemical potential $\mu_0$  which is small enough that the partition function $\Xi_{\rm S}(\mu_0,\CC)$ can be estimated directly from a grand canonical simulation of the small particles: it is equal to the reciprocal of the probability that $n=0$.

Next we perform an MC simulation with $M$ MCS during which the small particle chemical potential increases from $\mu_0$ to $\mu_{\rm S}$.  Let the small-particle chemical potential after $t$ MCS be $\mu(t)$, and the number of small particles at this time be $n(t)$.   Then define
\beq
{\cal I}(\CC) = \int_0^M n(t) \frac{\mathrm{d}}{\mathrm{d}t} \mu(t) \, \mathrm{d}t \; .
\label{equ:def-I-int}
\eeq
The parameter $M$ and the dependence of $\mu$ on $t$ are arbitrary at this stage and can be chosen to optimise the accuracy of the method.  In our simulations, $\mu$ increases stepwise.  Let $\Delta \mu_j$ be the change in $\mu$ on the $j$th step and $n_j$ be the number of small particles in the system when that step takes place.  Then 
\beq
{\cal I}(\CC) = \sum_{j=1}^K n_j \Delta \mu_j 
\label{equ:def-I-step}
\eeq
where $K$ is the number of steps.  (A simple choice is to make one step per MCS so that $M=K$, but other choices are possible.)
A consistent choice of the weight is then $\hat{W}(\CC) = \Xi_{\rm S}(\mu_0,\CC) \ee^{\beta {\cal I}(\CC) + \beta U_{\rm c}(\CC) }$.    Then Crooks' fluctuation formula~\cite{Crooks2000} (analogous to Jarzynski's equality~\cite{Jarzynski1997}) can be used to show that (\ref{equ:jarz-result}) holds.  A short derivation is given in Appendix~\ref{app:jarz}, see also Refs.~\onlinecite{Hummer2001,Neal2001,Oberhofer2009}. 
Alternatively we may repeat the same computation $m$ times (always with the same $\CC$) and take
\beq
\hat{W}(\CC) = \Xi_{\rm S}(\mu_0,\CC) \ee^{ \beta U_{\rm c}(\CC)} \frac{1}{m} \sum_{p=1}^m \ee^{\beta {\cal I}_p(\CC) }
\label{equ:Wm}
\eeq
where ${\cal I}_p(\CC)$ is the value of (\ref{equ:def-I-step}) obtained on the $p$th iteration of the computation.  Finally, using (\ref{equ:Wm}) in (\ref{equ:ratio}) gives the weight factors $\hat{w}_i$.

We summarise the parameters of the TL method.  The numbers of configurations used for the coarse and fine estimates are $\NNc,\NNf$.  These should both be large in order to have accurate results.  In addition, the Jaryznski integration depends on $M$ which is the total number of MCS in a Jarzysnki run, and $m$ which is the number of Jarzynski runs for each $\CC$.  The computation of $\hat{\Delta}$ is the dominant computational cost in the method, which scales as $mM\NNf$.  The choice of these parameters is discussed in Sec.~\ref{sec:convergence}.  The other choice that must be specified is the annealing schedule: that is, the dependence of $\mu$ on $t$ in (\ref{equ:def-I-int}), as well as the initial chemical potential $\mu_0$ and the number of steps $K$.  The choice of this schedule is discussed in Appendix~\ref{app:sim_details}.

\section{Numerical results}
\label{sec:results}

\subsection{Critical behaviour}

As anticipated in Sec.~\ref{sec:obs}, we focus on the probability distribution of $N$, the number of large particles in the system.  We analyse this distribution based on the theory of liquid-vapour critical points, but neglecting field-mixing effects~\cite{wilding1995critical}.  At this level of theory, 
the distribution of $N$ has a characteristic (universal) form at the critical point, with two peaks of equal height, separated by a trough.   The ratio of probability between the peak and trough is 0.46, in contrast to the phase-coexistence regime for which this ratio scales with system size as $\ee^{-\kappa L^2}$, where $\kappa$ is proportional to the surface tension.  The two peaks correspond to particle densities whose difference scales as $L^{-\beta/\nu}$  where $\beta,\nu$ are  the critical exponents for the  3d Ising universality class, so that $\beta/\nu\approx0.52$, see for example~Ref.~\onlinecite{wilding1995critical}. Hence the variance of the distribution decreases as $L^{-2\beta/\nu}$, as expected at liquid-vapour critical point. (These critical fluctuations are much larger than a comparable three-dimensional system with short-ranged correlations, where the variance decreases as $L^{-3}$).

In order to characterise the shape of the distribution, we define a scaling variable $X$ associated with the number of large particles by subtracting the mean of $N$ and dividing by the standard deviation:
\beq
 \delta N  = N - \langle N \rangle, \qquad X  = \frac{ \delta N  }{  \left\langle \delta N^2 \right\rangle^{1/2} } \; .
\eeq
The probability density for this variable is denoted by $P_L(X)$, where $L$ indicates the system size.

\subsection{Size ratio $q=\frac14$}

For $q=(1/4)$, direct simulation of the FG system is challenging, especially in the grand-canonical ensemble.  However, the TL method is applicable.  
Fig.~\ref{fig:ratio4} shows results obtained using this method, in a system of size $L=5\sigB$.  
Fig.~\ref{fig:ratio4}(a) shows that the critical point in this system occurs at $\etaS \approx 0.4080$.  Taking this value and making an appropriate choice of $\muB$, the distribution of $N$ matches well to the universal critical distribution. 

Fig.~\ref{fig:ratio4}(b) shows results at $\etaS \approx 0.4025$, for both coarse and fine models.  One sees that the coarse model is very close to its critical point, but the corresponding distribution for the fine model does not match the critical form.  In the fine system, the bimodality of the distribution is less pronounced than the critical form, indicating that this system is in the one-phase regime, for this value of $\etaS$.  The physical interpretation of this result is that three-body (and higher) interactions in the fine system act to reduce the attractive forces, which pushes the critical point to a larger value of $\etaS$.  This shift is $\Delta\etaS \approx 0.006$ for these parameters.  This is a small effect but our method is accurate enough to resolve it.  The qualitative difference between the FG distributions in Figs.~\ref{fig:ratio4}(a,b) emphasises that critical properties are very sensitive to model parameters: the values of $\etaS$ differ by less than $2\%$.

Note also that estimating the critical point by this method is only accurate up to an error that decays as a power law in $L$.  This error has contributions from field-mixing effects and from corrections to scaling~\cite{wilding1995critical}.
For the systems considered here, this correction might be comparable with $\Delta\etaS$ itself.  However,  we expect a similar correction for both coarse and fine systems, in which case our measured $\Delta\etaS$ should be a reasonable estimate of the difference in the critical points between coarse and fine systems.

For the system shown in Fig.~\ref{fig:ratio4}(a), a histogram of the reweighting factors $\hat{w}_i$  is shown in  Fig.~\ref{fig:ratio4}(c).  If the coarse and fine models matched exactly and the Jarzynski integration was perfect then all weight factors would be equal to unity.   We see that some weights are significantly larger than $1$: these are predominately coming from configurations whose probabilities are underestimated by the coarse model.  Such configurations must be given larger weight when estimating observable quantities for the fine model, and this is the effect of the correction $\hat\Delta$.

\begin{figure}
\includegraphics[width=80mm]{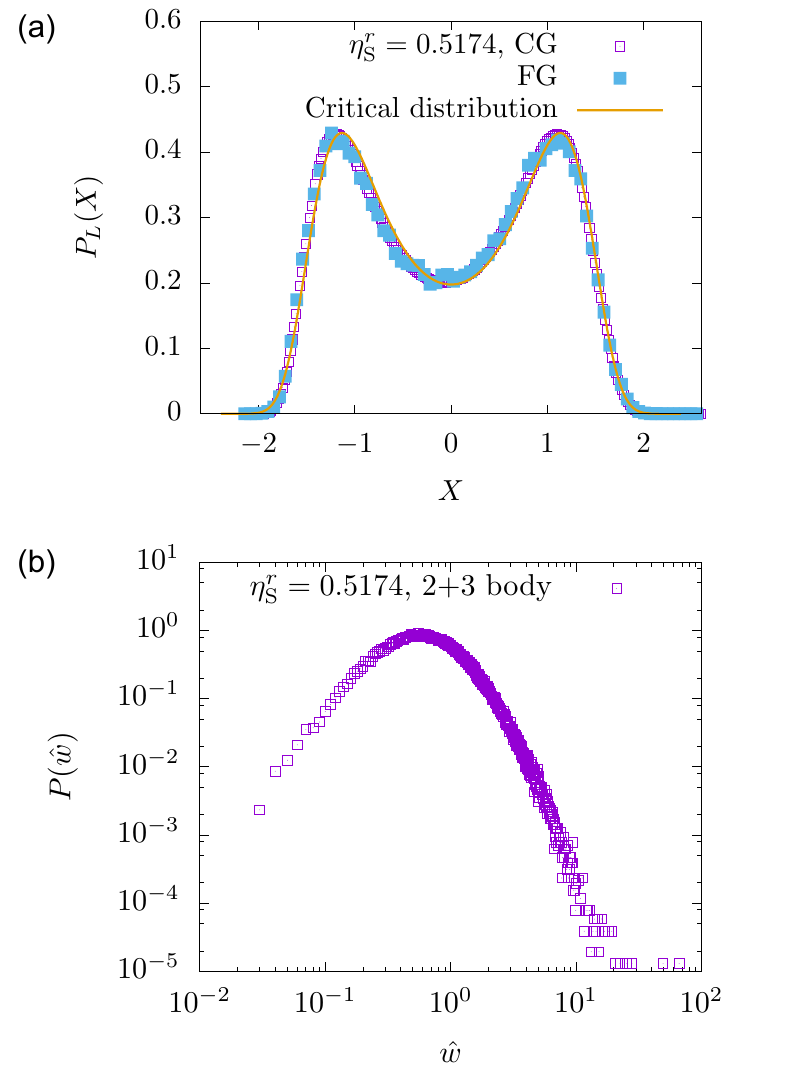}
\caption{Results for $q=(2/5)$ and $L=6.4\sigB$, using the TL method with the (2+3)-body coarse model.
(a) Histogram of the large-particle density in the fine system (FG) at $\etaS = 0.5174$, compared with the (universal) critical distribution and the corresponding distribution for the (2+3)-body coarse model (CG).  
All three distributions match very closely, indicating that the coarse model is an almost-exact representation of the fine model.
(b) Histogram of reweighting factors.  The distribution is peaked near $\hat{w}_i=1$, which confirms that the (2+3)-body coarse model is an accurate description of the fine one.
}
\label{fig:ratio25}
\end{figure}

\subsection{Size ratio $q=\frac25$}
\label{sec:q25}

We now turn to a size ratio $q=(2/5)$.   This system was analysed by direct simulation in Ref.~\onlinecite{Loverso2006}, using the grand-canonical ensemble, with cluster moves~\cite{Vink2004} that remove small particles at the same time as inserting large ones (and vice versa).
As in this work, the critical point was estimated by comparison with the universal order-parameter distribution: they found  $\etaS=0.5215\pm0.0001$, in a system of size $L=10\sigB$.  We note that the TL method does not outperform these previous methods for this moderate value of $q$: the TL method is designed to be effective for smaller $q$, which is the most challenging regime.  For example,  the next section considers $q=(2/13)$.
 
 Ref.~\onlinecite{Loverso2006} also shows that three-body interactions are significant for the size ratio $q=(2/5)$, in that the critical points differ significantly between the fine model and the coarse model with energy (\ref{equ:Ucoarse}).
 We show in Appendix~\ref{app:coarse-three} that the differences between the coarse and fine systems are too large for the TL method to be able to correct the coarse-graining error in that case
 We also explain how this
problem can be avoided.  Specifically, we define a new coarse model whose energy is
\beq
\beta U_{\rm 2,3}(\CC) =  \beta U_{\rm c}(\CC) +  \beta U_{\rm 3}(\CC)
\label{equ:U23}
\eeq
with $U_{\rm c}$ given by (\ref{equ:Ucoarse}) and 
\beq
U_{3}(\CC) =\sum_{i=1}^{N-2} \sum_{j=i+1}^{N-1} \sum_{k=j+1}^N V_3(R_{ij},R_{ik},R_{jk})
\label{equ:U3}
\eeq
where $R_{ab}=|\bm{R}_a-\bm{R}_b|$ is the distance between particles $a$ and $b$.  That is, $V_3$ is a 3-body interaction potential which we have chosen to represent as a function of the three relevant interparticle distances [see Appendix~\ref{app:coarse-three} and recall also (\ref{equ:pm-pc})].  

We refer to the model with energy (\ref{equ:U23}) as the (2+3)-body coarse model.
The three-body interaction is repulsive: it weakens the effective attractive attraction between particles so the critical point for the (2+3)-body coarse model is at a higher volume fraction $\etaS$ than the critical point of the two-body coarse model (\ref{equ:Ucoarse}).

Results using the (2+3)-body coarse model within the TL method are shown in Fig.~\ref{fig:ratio25}.
In particular, Fig.~\ref{fig:ratio25}(a) shows histograms of the large-particle density for coarse and fine models at $\etaS=0.5174$, compared with the critical form.  Two results are notable: first, the fine model matches well the critical distribution, so we conclude that this system is close to its critical point.  Second, the fine model matches extremely accurately the coarse model.  This indicates that the (2+3)-body coarse model is an almost-quantitative description of the fine system.  This is confirmed by Fig.~\ref{fig:ratio25}(b) which shows the distribution of the $\hat{w}_i$ which is peaked near $\hat{w}_i=1$, consistent with the fact that the coarse-grained model is very accurate.  Compared with Ref.~\onlinecite{Loverso2006}, the difference (less than $1\%$) in the estimated critical value of $\etaS$ is presumably attributable to the smaller system size $L=6.4\sigB$ used in this work.

Since the (2+3)-body coarse model gives an almost-exact representation of the critical point, we conclude that four-body and higher interactions are negligible at this state point.  This is perhaps not surprising since the (average) volume fraction of the large particles is around $0.22$ at the critical point, and the effects of four-body and higher interactions are suppressed as the fourth power of the density.  However, the result is promising in that it shows how low-order corrections to two-body interaction potentials can already give a significant improvement in accuracy.

\begin{figure}
\includegraphics[width=80mm]{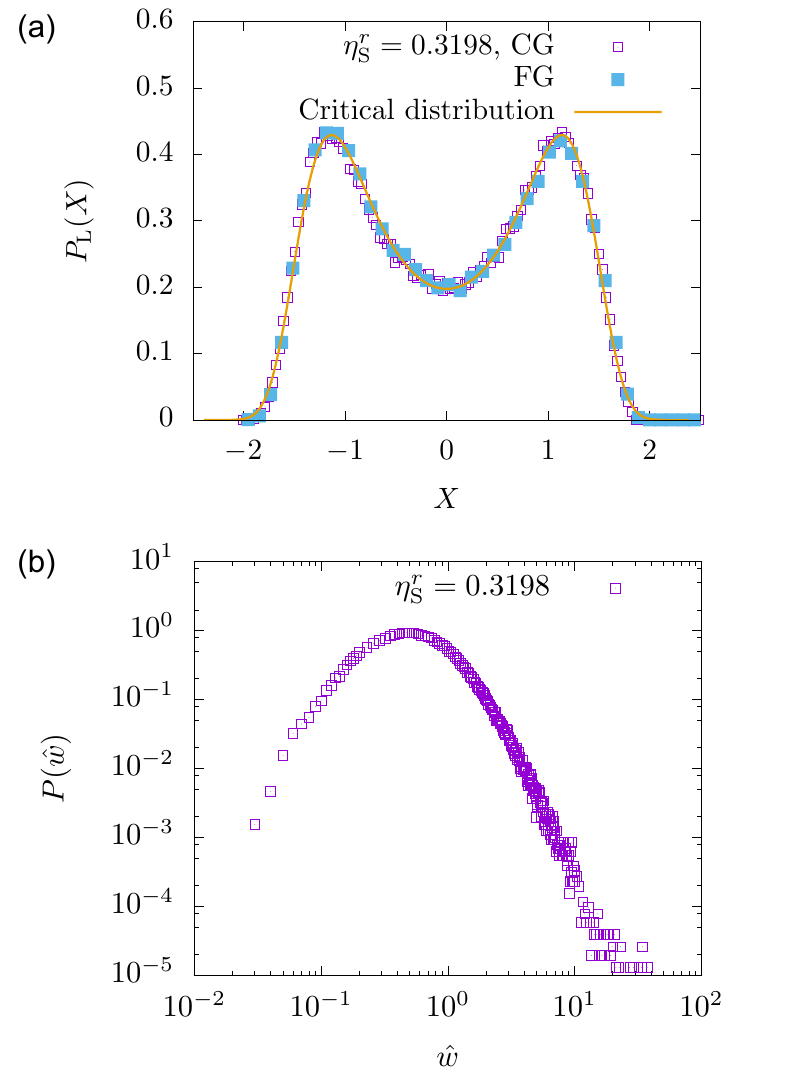}
\caption{Results for $q=(2/13)$ and $L=4.8\sigB$.  (a) Distributions of the number of large particles for coarse and fine models, compared with the universal critical distribution.  (b) Histogram of weights.
For this value of $q$ then the coarse model is perfect so all of the variance of the weights comes from fluctuations in the Jarzynski integration.
}
\label{fig:ratio65}
\end{figure}

\subsection{Size ratio $q=\frac{2}{13}$}

As a third example system, we consider  $q=(2/13)\approx 0.1538$ for which the two-body coarse model is exact~\cite{Binder2014}.   Results are shown in Fig.~\ref{fig:ratio65}.  They show that the FG and CG models do indeed have the same order-parameter distribution, as they must.  There are many small particles (recall Fig.~\ref{fig:snapshot-phase}, which has over 20,000 of them)  so direct simulation of this system would be extremely challenging.  

Within the TL method it is known that $\langle \hat\Delta\rangle=0$ because the coarse model is exact.  However, it is not easy to confirm this result numerically because the Jarzynski integration requires significant computational cost.  Nevertheless,  Fig.~\ref{fig:ratio65}(b) shows that the distribution of weights $\hat{w}_i$ is peaked near unity, which means that the method is effective, even for this challenging test case.   Fig.~\ref{fig:ratio65}(a) confirms that the results are accurate.  The width of the distribution of weights comes entirely from the Jarzynski integration.  It is suppressed by choosing a large value of $M$, corresponding to a significant computational effort for the computation of each individual weight.  The small error on $P(N)$ is ensured by combining a sufficiently narrow distribution of weights (large $M$), with an average over many of these weights (sufficiently large $\NNf$).  For details of the simulations see Appendix~\ref{app:sim_details}.  The scaling of the numerical error with respect to the algorithmic parameters is the subject of the next section.

The physical insight of Fig.~\ref{fig:ratio65} is that the TL method is applicable even when the number of small particles is very large, recall Fig.~\ref{fig:snapshot-phase}(b).

\section{Error estimation and choice of algorithmic parameters}
\label{sec:convergence}

\subsection{Overview}
\label{sec:con-overview}

This section analyses the convergence shown in (\ref{equ:A-conv}).  
We will show that as $\NNc,\NNf\to\infty$ then
\beq
\big\langle \hat{A} - \langle A \rangle_{\rm f}\big\rangle \to 0
\label{equ:mean-ok}
\eeq
where the outer average (without subscript) is over all random aspects of the TL algorithm, which includes the generation of the original $\NNc$ coarse samples, the selection of the $\NNf$ samples to be used in (\ref{equ:hat-Delta}), and the Jarzynski integration.  The mean square error of the method is
\beq
h^2  = \left\langle \left( \hat{A} - \langle A \rangle_{\rm f}\right)^2 \right\rangle   \; .
\label{equ:var-h}
\eeq
We will analyse the scaling of the error $h^2$ in the limit $\NNc,\NNf\to\infty$.

Our analysis requires several assumptions, which are all satisfied in the physical situations of interest.  Note first that the weighting factor $w(\CC)=p_{\rm m}(\CC)/p_{\rm c}(\CC)$ may be very large if $p_{\rm c}$ is very small.  This happens if there are configurations of the FG model that are very rare in the CG model.  We assume that the variance of $w(\CC_j)$ is finite, which requires that the CG model is not too inaccurate.  In fact, it is useful to quantify the accuracy of the CG model  
in terms of the $\chi^2$-divergence~\cite{Chen2005,Agapiou2017} between  $p_{\rm c}$ and $p_{\rm m}$, which  is defined as
\beq
D_{\chi^2}( p_{\rm m} \| p_{\rm c} ) = \int \left( \frac{p_{\rm m}(\CC)}{p_{\rm c}(\CC)} - 1 \right)^2  \cdot p_{\rm c}(\CC) \mathrm{d}\CC
\label{equ:Dchi2}
\eeq
This quantity is non-negative in general and equals zero only if the coarse model is perfect, that is $p_{\rm c}=p_{\rm m}$.  Smaller values of $D_{\chi^2}$ correspond to coarse models that are more accurate.  Our assumption is that
\beq
D_{\chi^2}( p_{\rm m} \| p_{\rm c} ) < \infty \; .
\label{equ:Dchi2-finite}
\eeq
We also require an assumption on the variance of the weight factors $\hat{W}$ generated by the Jarzynski integration method.  
As a minimal assumption, we suppose that this variance is finite, which means that 
\beq
\big\langle \hat{W}(\CC_j)^2  \big\rangle  < \infty \; .
\label{equ:OmTot}
\eeq
The expected behaviour of the weights $\hat{W}$ is discussed below, but we expect that (\ref{equ:OmTot}) holds in all situations of physical interest.

In addition to (\ref{equ:Dchi2-finite},\ref{equ:OmTot}), we make several other assumptions, to simplify the analysis.  
In particular, we assume that all the samples (both coarse and fine) are independent of each other.  In practical situations this assumption will not hold exactly, but we argue that our results still have the correct scaling with respect to the algorithmic parameters.  For simplicity, we also assume throughout that $A$ is a bounded quantity, so that  
\beq
|A(\CC)-\langle A\rangle_{\rm f}| \leq a
\label{equ:bounded-A}
\eeq 
for some $a$, independent of $\CC$.

\subsection{Convergence of the mean}

To establish (\ref{equ:mean-ok}), we first note from (\ref{equ:hat-Ac}) that 
\beq
\big\langle \hat{A}_{\rm c} - \langle A \rangle_{\rm c}\big\rangle = 0
\label{equ:Ac-mean}
\eeq 
which follows by linearity of the average, because every $\CC_i$ is distributed as $p_{\rm c}$.
Similarly,
\beq 
\left\langle A(\CC_j) \frac{p_{\rm m}(\CC_j) }{ p_{\rm c}(\CC_j) }  \right\rangle = \langle A\rangle_{\rm f}  \; .
\label{equ:Amc}
\eeq

The next step is to quantify the approximation in (\ref{equ:mean-w}).  
To this end define $\xi=\Xi_{\rm f}/\Xi_{\rm c}=\langle \hat{W}(\CC_j)\rangle$ and also
\beq
\epsilon = \frac{1}{\NNf}\sum_j [\hat{W}(\CC_j)/\xi]-1\; ,
\label{equ:def-eps}
\eeq
Clearly $\langle\epsilon\rangle=0$, and $\eps$ is an average of independent mean-zero variables.  Also $\hat{W}(\CC_j)$ has a finite variance by (\ref{equ:OmTot}), so that $\eps$ obeys a central limit theorem, and its typical value is therefore small as  $\NNf\to\infty$.
Also from (\ref{equ:ratio})
\beq
\hat{w}(\CC_i) = \frac{\hat{W}(\CC_i)}{\xi} \left[ 1 - \frac{\eps}{1+\eps} \right] 
\label{equ:w-eps}
\eeq
which is a more precise statement of (\ref{equ:mean-w}).
Observe that  $\sum_j \hat{w}(\CC_j)=1$ so (\ref{equ:hat-Delta}) is identical to 
\beq
\hat{\Delta} = \frac{1}{\NNf} \sum_j (   A(\CC_j) - \langle A \rangle_{\rm f}  ) ( \hat{w}(\CC_j) -1  ) \; .
\label{equ:Delta-centre}
\eeq
Hence, using again that the $\CC_j$ are distributed as $p_{\rm c}$, one has
\beq
\langle \hat\Delta \rangle = \langle A \rangle_{\rm f} - \langle A \rangle_{\rm  c}
+ \langle \hat{\mathcal{E}} \rangle
\label{equ:Delta-E}
\eeq
with
\beq
\hat{\mathcal{E}} = \frac{1}{\NNf} \sum_j (  A(\CC_j) -  \langle A \rangle_{\rm f}   ) \hat{w}(\CC_j) \;.
\label{equ:E-hat-def}
\eeq
Using (\ref{equ:w-eps}) and that $\eps$ is small, we can establish a bound on the average of $\hat{\mathcal E}$. Details are given in Appendix~\ref{app:eps}.
The result is
\beq
\big\langle \hat{\Delta} \big\rangle  = \langle A \rangle_{\rm f} - \langle A \rangle_{\rm c} + O(\NNf^{-1}) \; .
\label{equ:mean-Delta}
\eeq
With (\ref{equ:Acf},\ref{equ:Ac-mean}), this establishes (\ref{equ:mean-ok}).  

We have emphasised throughout that the Jarzynski method for computing weights is useful because it minimises systematic errors.  To be specific, the simplest alternative to this approach would be to  use a standard thermodynamic integration to estimate $p_{\rm m}/p_{\rm c}$, with the same computational effort $M$ as was used in the Jarzynski method.  The disadvantage of this approach is that one expects a systematic error on $\hat\Delta$ from the finite rate of integration, which would appear in (\ref{equ:mean-Delta}) as a correction at ${\cal O}(M^{-1})$.  (Recall that $M$ is the total number of MCS used in the Jarzynski integration.)  In this case convergence of the method requires a joint limit $\NNc,\NNf,M\to\infty$, instead of the (simpler) limit $\NNc,\NNf\to\infty$ required here.

\subsection{Mean square error}
\label{sec:error}

This section estimates the 
the error defined in (\ref{equ:var-h}), and how it scales with the parameters $\NNc,\NNf,M$.  
Let
\beq
h_{\rm c}^2  
= \left\langle \left( \hat{A}_{\rm c} - \langle A \rangle_{\rm c}\right)^2 \right\rangle   \; .
\eeq
Assuming that the $\NNc$ samples used in $\hat{A}_{\rm c}$ are independent,
it is immediate that 
\beq
h_{\rm c}^2 = \frac{\langle A^2 \rangle_{\rm c} - \langle A \rangle_{\rm c}^2 }{\NNc}
\label{equ:hc-result}
\eeq
Using (\ref{equ:bounded-A}), the numerator is bounded above by $a^2$, so $h_{\rm c}^2\to0$ as $\NNc\to\infty$.
In practice, (\ref{equ:Ac}) uses a sequence of samples from a MC simulation, so that samples which are nearby in the sequence are not independent.  However, these correlations will be short-ranged, as long as the MC simulation is converged.  In this case $h_{\rm c}^2$ is larger than (\ref{equ:hc-result}) by a constant factor but the scaling with $\NNc$ is the same.

For the $\NNf$ samples used in (\ref{equ:hat-Delta}),  it is again convenient to assume that these samples are independent of each other and of the coarse samples.
The mean square error of $\hat\Delta$ is
\beq
h_{\rm f}^2  =  \left\langle \left( \hat{\Delta} -  \langle A \rangle_{\rm f} + \langle A \rangle_{\rm c}\right)^2 \right\rangle   \; .
\eeq
Writing $ \hat{\Delta} = (\hat{\Delta} - \langle\hat\Delta\rangle) + \langle\hat\Delta\rangle $ and using (\ref{equ:mean-Delta}) yields 
\beq 
h_{\rm f}^2  =  \left\langle \left( \hat{\Delta} -  \langle \hat\Delta\rangle \right)^2 \right\rangle + {\cal O}(\NNf^{-2}) \; .
\eeq
We show below that $h_{\rm f}^2 = O(\NNf^{-1})$ so it is consistent to drop the correction term.  Using (\ref{equ:Delta-centre})  we then obtain the leading order result
\beq
h_{\rm f}^2 = \mathrm{Var}\left(\frac{1}{\NNf}\sum _j ( A(\CC_j)-\langle A \rangle_{\rm f})  (\hat{w}(\CC_j)-1) \right) . 
\label{equ:hf-var}
\eeq
Substituting for $\hat{w}$ with (\ref{equ:w-eps}), the leading-order behaviour can be obtained by dropping all terms proportional to $\eps$.  Hence
\beq
h_{\rm f}^2 = \frac{1}{\NNf} \mathrm{Var} \left[ ( A(\CC_j)-\langle A \rangle_{\rm f})  \left(\hat{W}(\CC_j)/\xi-1\right) \right] \; .
\label{equ:hf-unpleasant-ii}
\eeq
It follows from (\ref{equ:bounded-A}) that for any random variable $B$ then $\mathrm{Var}[( A(\CC_j)-\langle A \rangle_{\rm f})B(\CC_j)]\leq a^2\langle  B(\CC_j)^2\rangle$.  Hence (\ref{equ:hf-unpleasant-ii}) reduces to
\beq
h_{\rm f}^2 \leq \frac{a^2}{\NNf} \left \langle \left([\hat{W}(\CC_j)/\xi]-1\right)^2 \right \rangle.
\label{equ:hf-simplified}
\eeq

The physical interpretation of (\ref{equ:hf-simplified}) is that the scaling of the error is controlled by the variance of the weights $\hat{W}$.  
There are two contributing factors to this variance, which are the randomness in the Jarzynski integration and the fact that $\CC_i$ is distributed as $p_{\rm c}$.

We split the error estimate into two pieces, one from each source of randomness.  Assume first
that the reweighting factors $\hat w$ are obtained exactly so $\hat{w}_i = p_{\rm m}(\CC_i)/p_{\rm c}(\CC_i)$.  In this case all randomness comes from the $\CC_i$.
{Hence (\ref{equ:hf-simplified}) reduces to}
\beq
h_{\rm f}^2 \leq h_{{\rm f},0}^2, \qquad h_{{\rm f},0}^2 = \frac{a^2}{\NNf} D_{\chi^2}( p_{\rm m} \| p_{\rm c} ) \; .
\label{equ:hf0}
\eeq
The case where $p_{\rm c}=p_{\rm m}$ is instructive, it corresponds to a perfect coarse model so that $D_{\chi^2}=0$ and $h_{\rm f,0}^2=0$.  Note however that the inequality in~(\ref{equ:hf0}) is valid only if the weights $\hat{w}$ are exact.

To account for the randomness in these weights, recall that the average of $\hat{W}(\CC)$ obeys (\ref{equ:jarz-result}), 
but (\ref{equ:hf-unpleasant-ii}) also depends on the fluctuations of this quantity.
To estimate the variance of $\hat{W}(\CC)$, note that ${\cal I}$ in (\ref{equ:def-I-step}) is a sum of weakly correlated increments, so one expects its statistics to be ruled by a central limit theorem~\cite{Neal2001}.
For this reason, we assume that for any fixed $\CC$, the sum ${\cal I}$ has a Gaussian distribution with variance $\Lambda/M$.
We assume that $\Lambda$ is independent of $\CC$, for convenience.

Recall from (\ref{equ:Wm}) that we obtain the weight $\hat{W}$ by averaging over $m$ independent realisations of the Jarzynski integration.  If $m=1$ one sees that
$\log\hat{W}(\CC)$ differs from ${\cal I}$ by an additive constant and is therefore Gaussian with variance $\Lambda/M$. 
It follows that 
$
\langle \hat{W}(\CC)^2 \rangle_{\rm J}  = \langle {W}(\CC) \rangle_{\rm J}^2 \exp(\Lambda/M) 
$.
For $m>1$ the variance of $\hat{W}$ is suppressed by a factor of $m$, which leads to
\beq
\langle \hat{W}(\CC)^2 \rangle_{\rm J}  -\langle \hat{W}(\CC) \rangle_{\rm J}^2 =  \frac{\langle \hat{W}(\CC) \rangle_{\rm J}^2}{m} [ \exp(\Lambda/M)  - 1 ] \; .
\eeq
Hence
\begin{multline}
 \big\langle \big([\hat{W}(\CC)/\xi]-1\big)^2 \big\rangle_{\rm J} 
= 
 \left(\frac{p_{\rm m}(\CC)}{p_{\rm c}(\CC)} - 1\right)^2
\\ + 
\left(\frac{p_{\rm m}(\CC)}{p_{\rm c}(\CC)}\right)^2 \frac{ \exp(\Lambda/M) - 1 }{m}
\; .
\end{multline}
%
{Using this result in (\ref{equ:hf-simplified}) yields}
\beq
h_{\rm f}^2 \leq h_{{\rm f},0}^2 +  h_{{\rm f,J}}^2
\label{equ:hftot}
\eeq
with
\beq
h_{{\rm f,J}}^2 =   \frac{a^2}{\NNf} \left[ D_{\chi^2}( p_{\rm m} \| p_{\rm c} ) + 1 \right] \frac{ \exp(\Lambda/M) - 1 }{m}
\label{equ:hfJ}
\eeq
The key point here is that even if the coarse model is perfect ($D_{\chi^2}=0$), there is randomness in the Jarzynski integration which generates statistical errors.  From (\ref{equ:hf0},\ref{equ:hftot},\ref{equ:hfJ}) one sees that indeed $h_{\rm f}^2=O(\NNf^{-1})$, as asserted above.

Combining all these ingredients and recalling our various assumptions, we find the scaling of the mean square error,
\beq
h^2 \leq a^2 \left[ \frac{1}{\NNc} +  \frac{1}{\NNf} H_{\rm f} \right]
\label{equ:h2final}
\eeq
with 
\beq 
H_{\rm f} = D_{\chi^2}( p_{\rm m} \| p_{\rm c} ) + \left[ D_{\chi^2}( p_{\rm m} \| p_{\rm c} ) + 1 \right] \frac{ \exp(\Lambda/M) - 1 }{m} \; .
\label{equ:Hf}
\eeq

\subsection{Implications for optimisation of the TL method}
\label{sec:tl-opt}

The results (\ref{equ:h2final},\ref{equ:Hf}) establish useful properties of the TL algorithm. 
The computational effort for the Jarzynski integration is proportional to the product $mM\NNf$.  We infer  that the optimal parameter choice is to take $\Lambda/M=O(1)$ in order that the exponential factor in (\ref{equ:Hf}) is not too large, and then to devote the remaining computational effort towards larger $\NNf$.   The benefits of increasing $m$ are less than those of increasing $\NNf$ or $M$, so the choice $m=1$ is likely optimal.  However, it may be convenient in practice to take some other small integer since (for example) increasing $m$ can be achieved by additional parallel processing while increasing $M$ requires longer simulation (wall) time.  The parameters of the Jarzynski integration (particularly the annealing schedule) should be chosen to minimise $\Lambda$ (at fixed $M$).     It is obviously desirable that $D_{\chi^2}$ be small  -- this simply means that the coarse model should be as accurate as possible.  In fact, $D_{\chi^2}$ can be related to the exponential of a free energy difference between $p_{\rm m}$ and $p_{\rm c}$, via (\ref{equ:Dchi2}).  Hence  we expect that this quantity will scale exponentially in the system size -- very accurate coarse models are required if the method is to be effective in very large systems.  

The factors of $D_{\chi^2}$ and $\ee^{\Lambda/M}$ in (\ref{equ:Hf}) both indicate that the method can suffer from exponentially growing errors.  However, the numerical results of Sec.~\ref{sec:results} show that the method is still practical in realistic soft-matter systems.  The fact that $D_{\chi^2}$ and $\Lambda$ may be large in practice indicate that the majority of the computational effort in this method should be spent on the computation of $\hat\Delta$, which requires that both $M$ and $\NNf$ should be large.  By contrast, we found that the error $h_{\rm c}$ coming from $\NNc$ tends to be small, in the systems analysed here.

\section{Conclusion}
\label{sec:conc}

We have presented a two-level method for correction of coarse-graining errors and we have demonstrated it using  the AO mixture.  We have found that three-body interactions shift the critical point to higher $\etaS$, for $q=(2/5)$ and $q=(1/4)$.   For $q=(2/5)$ we characterised the three-body interactions, which have a significant effect on the critical point, and also we showed that four-body and higher interactions are negligible, at criticality.

We also analysed the errors associated with the method, in Sec.~\ref{sec:convergence}.  The first important result of that section is that the method is accurate as $\NNc,\NNf\to\infty$ but does not require any limit of large $M$.  That is, we do not require that the individual weight factors $\hat{w}_i$ are accurate measures of $p_{\rm m}/p_{\rm c}$.  Instead we require only that these weights have the right average, which is ensured by (\ref{equ:mean-w}).  The results (\ref{equ:h2final},\ref{equ:Hf}) show how the parameters of the method should be chosen, as discussed in Sec.~\ref{sec:tl-opt}.

As noted above, the general two-level method is similar in spirit to free-energy perturbation theory and importance sampling.   As such, it has a range of potential applications.  Its main strength is that direct simulation of the whole system is never required.  For example, in the system presented here, the large particles remain fixed while the small particles are introduced, which avoids the difficulty of moving (or inserting) large particles in a system like Fig.~\ref{fig:snapshot-phase}.  If one attempted direct MC simulation of the full system then the error (\ref{equ:h2final}) would be inversely proportional to the total number of MCS, but it would be proportional to the number of MCS required to generate an independent sample of the full system.  This number is expected to be extremely large in practice.  (In particular, it would likely be much larger than $\Lambda$.)  On the other hand, direct MC simulation does not suffer from problems associated with large values of $D_{\chi^2}$, which can hamper the TL method.  That is, the TL method requires rather accurate coarse descriptions in order to be successful (see also Fig.~\ref{fig:ratio25-bad} in Appendix~\ref{app:coarse-three}).  Nevertheless, we have shown that information from Jarzynski integration can be used to learn about corrections to coarse models, including three-body interactions.  Combination of the TL method with machine-learning methods~\cite{Behler2007,Bartok2010,Cheng2019} seems to be a promising future direction.

Of course, there are many strategies for direct MC simulation.  In particular there are tailored algorithms for mixtures of large and small particles, including the cluster moves used in Ref.~\onlinecite{Loverso2006}, other cluster moves~\cite{Liu2006}, and staged (grand-canonical) particle insertion~\cite{Ashton2011}.   These methods reduce the number of MCS required to generate an independent sample.   
\rlj{For the specific case of the AO model, the method of Ref.~\onlinecite{Dijkstra2006} is also available.}  
A direct comparison with these methods is tricky because the computational efficiency depends strongly on details of the implementation.  However, the results shown here are promising, especially given the relatively high volume fractions of the small particles.  We also emphasise that the TL method is generic, while these other algorithms have been tailored to the specific problems of interest.  

Another strength of the TL method is that while the computational effort of the Jarzynski integration is proportional to $mM\NNf$, this effort is easily split into $m\NNf$ independent simulations, each involving $M$ MCS.   This means that the method is well-suited to parallel computation, in contrast to direct MC simulation where very long individual simulations may be required.

In the future, we intend to test the TL method on mixtures of large and small hard spheres, \rlj{for which accurate coarse models can be computed theoretically~\cite{Dijkstra1998,Dijkstra1999-pre,Ashton2011depletion}. }   The method will allow analysis of phase behaviour in these systems, and also quantification of the effects of many-body interactions~\cite{Ashton2014-jcp}.  Similar questions arises in systems where the large particles are not spherical, in which case the depletion interaction has a more complicated character~\cite{Odriozola2008,vanAnders2014,Law2016} and can lead to new phase behaviour~\cite{vanAnders2014,Ashton2015}. There are also potential applications of the TL method to other soft matter systems including star polymers and dendrimers, for which accurate coarse-grained models are available~\cite{Jusufi1999,Likos2001,Marzi2012}, and whose phase behaviour is rich and complex~\cite{Lenz2012,Wilding2014}.

\begin{acknowledgments}
This work was supported by the Leverhulme Trust through research project grant number RPG--2017--203.
\end{acknowledgments}

\begin{appendix}

\section{Additional theoretical details}

\subsection{Grand canonical ensemble}
\label{app:gce}

To analyse these fluids in the grand canonical ensemble requires a definition of the integration measure in 
 (\ref{equ:Af}), which includes a sum over the numbers of particles and an integral over their positions.  We take
\begin{multline}
\langle A \rangle_{\rm f} = \frac{1}{\Xi_{\rm f}} \sum_{N=0}^\infty \sum_{n=0}^\infty \frac{ \ee^{\beta \muS n + \beta\muB N}}{n! N! \sigB^{3N}\sigS^{3n}} 
\int 
A({\cal C}) 
\ee^{-\beta U_{\rm f}(\CC,\FF)}
\\ \times 
\mathrm{d}\RR_1\dots \mathrm{d}\RR_N  
\mathrm{d}\rr_1\dots \mathrm{d}\rr_n \; ,
\end{multline}
where all particle positions are integrated over the box $[0,L]^3$.
This formula is consistent with (\ref{equ:pf-ao}), recall that $\CC=(N,\RR_1,\dots,\RR_N)$ and $\FF=(n,\rr_1,\dots,\rr_n)$ and note
that the zero of chemical potential has been chosen so that the thermal de Broglie wavelength does not appear, as usual in simulation studies.   Note also that the definition of the integration measure $\mathrm{d}\CC$ includes the factor of $\frac{1}{N!}$ that accounts for indistinguishability of particles, and similarly for $\mathrm{d}\FF$.

Also, normalisation of $p_{\rm f}$ requires
\begin{multline}
\Xi_{\rm f} = \sum_{N=0}^\infty \sum_{n=0}^\infty \frac{ \ee^{\beta \muS n + \beta\muB N}}{n! N! \sigB^{3N}\sigS^{3n}} 
\int 
\ee^{-\beta U_{\rm f}(\CC,\FF)}
\\ \times 
\mathrm{d}\RR_1\dots \mathrm{d}\RR_N  
\mathrm{d}\rr_1\dots \mathrm{d}\rr_n \; .
\end{multline}
Similar formulae hold for $\langle A \rangle_{\rm c}$ and $\Xi_{\rm c}$.

To compute $p_{\rm m}(\CC)$ from (\ref{equ:pm},\ref{equ:pf-ao}) one notes that the integrations over the particle positions can be performed independently and that each of them yields a factor of ${\cal V}_{\rm a}$, which is the volume accessible to a single small particle.  For configurations $\CC$ where no large particles are overlapping one finds 
\begin{align}
p_{\rm m}(\CC) &= \frac{\ee^{\beta\muB N}}{\Xi_{\rm f}}  \sum_{n=0}^\infty \frac{ \ee^{\beta \muS n}}{n!\sigS^{3n}} {\cal V}_{\rm a}^n
\end{align}
Recognising the sum as the Taylor expansion of an exponential and using the definition of $\etaS$ yields (\ref{equ:pm-ao}).

\subsection{Jarzynski average (\ref{equ:jarz-result})}
\label{app:jarz}

We outline the derivation of (\ref{equ:jarz-result}), following Ref.~\onlinecite{Crooks2000}.  As in (\ref{equ:def-I-step}), the chemical potential increases in a sequence of $K$ steps.  
Hence we consider a sequence of $K+1$ values of the  
chemical potential, $(\mu_0,\mu_1,\dots,\mu_{K})$, with $\mu_{K}=\muS$, the value of the chemical potential for the fine system.  Also $\Delta\mu_k = \mu_{k}-\mu_{k-1}$.  Between steps $k$ and $k+1$ one inserts and removes small particles using an MC algorithm that 
obeys detailed balance with respect to a probability distribution 
\beq
p_{k}(\CC,\FF) \propto \ee^{\beta\muB N + \beta\mu_k n - \beta U_{\rm f}(\CC,\FF)},
\eeq
which is similar to (\ref{equ:pf-ao}) with $\muS$ replaced by  $\mu_k$.  
To establish (\ref{equ:jarz-result}) from (\ref{equ:Wm}) it is sufficient to show that
\beq
\langle \ee^{\beta{\cal I}(\CC)} \rangle_{\rm J} = \frac{\Xi_{\rm S}(\muS,\CC)}{\Xi_{\rm S}(\mu_0,\CC)} \; ,
\label{equ:crooks}
\eeq 
where the average is with respect to this random MC algorithm and the small-particle partition function of (\ref{equ:XiS}) can be alternatively expressed as
\beq
\Xi_{\rm S}(\mu,\CC) = \int {\rm e}^{ \beta\mu n - \beta U_{\rm f}(\CC,\FF)} \mathrm{d}\FF \;.
\label{equ:XiS-part}
\eeq  

Let the sequence of small-particle configurations when the steps take place be $(\FF_1,\FF_2,\dots,\FF_K)$, and let the associated particle numbers be $(n_1,n_2,\dots,n_K)$.  The initial small-particle configuration 
is distributed as
\beq
\pi_0(\FF_1) = \frac{1}{\Xi_{\rm S}(\mu_0,\CC)} {\rm e}^{ \beta\mu_0 n_1 - \beta U_{\rm f}(\CC,\FF_1)}
\label{equ:pi0F0}
\eeq
which is a normalised probability density.
Then write
\begin{multline}
\langle \ee^{\beta{\cal I}(\CC)} \rangle_{\rm J} = \int  \pi_0(\FF_1) \left[ \prod_{k=1}^{K-1} q_k(\FF_{k}\to \FF_{k+1}) \right] 
\\ \times  \ee^{\sum_{k=1}^K \beta n_k (\mu_{k}-\mu_{k-1}) }  \mathrm{d}\FF_1 \mathrm{d}\FF_2 \dots \mathrm{d}\FF_K
\end{multline}
where  $q_k(\FF_{k}\to \FF_{k+1})$ is the probability (density) that the small particles evolve from $\FF_{k}$ to $\FF_{k+1}$, between steps $k$ and $k+1$.
Rearranging factors in the exponential term, this becomes
%
\begin{multline}
\langle \ee^{\beta{\cal I}(\CC)} \rangle_{\rm J} = \int  \left[ \prod_{k=1}^{K-1} q_k(\FF_{k}\to \FF_{k+1}) \ee^{\beta (n_{k}-n_{k+1})\mu_k } \right]
\\ \times \pi_0(\FF_1) \ee^{-\beta(n_1\mu_0 - n_K \muS)}  \mathrm{d}\FF_1 \mathrm{d}\FF_2 \dots \mathrm{d}\FF_K
\label{equ:big-prod}
\end{multline}
Detailed balance of the Markov chains implies that 
\begin{multline}
q_k(\FF_{k}\to \FF_{k+1}) \ee^{\beta (n_{k}-n_{k+1})\mu_k}=q_k(\FF_{k+1}\to \FF_{k}) 
\\ \times \exp\left[- \beta U_{\rm f}(\CC,\FF_{k+1}) + \beta U_{\rm f}(\CC,\FF_{k})\right]\;.
\end{multline}
Using this in (\ref{equ:big-prod}) together with (\ref{equ:pi0F0}) yields
\begin{multline}
\langle \ee^{\beta{\cal I}(\CC)} \rangle_{\rm J} = \frac{1}{\Xi_{\rm S}(\mu_0,\CC)} \int  \left[ \prod_{k=1}^{K-1} q_k(\FF_{k+1}\to \FF_{k}) \right] 
\\
\times \ee^{\beta \muS n_K - \beta U_{\rm f}(\CC,\FF_{K})}
\mathrm{d}\FF_1 \mathrm{d}\FF_2 \dots \mathrm{d}\FF_K
\end{multline}
The integrals can now be performed one by one (starting with $\FF_1$) and using that $q_k(\FF_{k+1}\to \FF_{k})$ is a normalised probability density for $\FF_{k}$.
The final integral over $\FF_K$ is of the form (\ref{equ:XiS-part}) with $\mu = \muS$ and one obtains (\ref{equ:crooks}).

\begin{figure*}
\includegraphics[width=175mm]{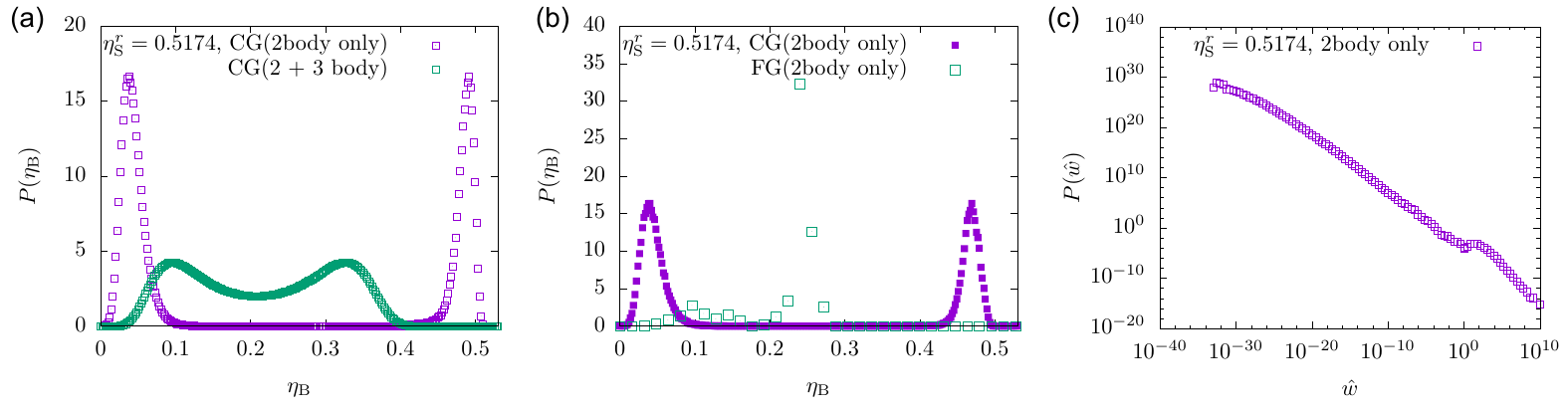}
\caption{Results for $q=(1/4)$ and $L=5\sigB$, at $\etaS = 0.5174$ showing how the TL method breaks down when the coarse model is not sufficiently accurate.  
(a) Histogram of the large-particle density for the 2-body coarse model and the (2+3)-body coarse model.
The coarse model with two-body interactions has two sharp peaks, showing phase coexistence.  The FG system is close to criticality, for these parameters, as demonstrated in Fig.~\ref{fig:ratio25}.
The coarse model with (2+3)-body interactions is close to its critical point.
(b) Estimate of the distribution for the number of large particles in the fine system obtained using (\ref{equ:Acf}), with the 2-body coarse model.  In this case the coarse model is not accurate and the estimated results for the fine system have large errors.  
(c) Corresponding distribution of reweighting factors.  Most samples have extremely small weights and a few have very large weights $\hat{w}_i\gtrsim 10^8$.  
}
\label{fig:ratio25-bad}
\end{figure*}

\section{Simulation details}
\label{app:sim_details}

For simulation of the coarse model we use grand-canonical MC simulations as in Ref.~\onlinecite{wilding1995critical}.  We restrict to particle insertion and removal (that is, no translational moves): this enables accurate results for fluid phases but helps to suppress crystallisation~\cite{Largo2006}.  We perform long MC runs (for example up to $10^8$ MCS, for which the computational effort is a few hours) and we extract $\NNf=128,000$ configurations for the computation of $\hat\Delta$.

The Jarzynski integration follows Sec.~\ref{sec:jarz-ao} and Appendix~\ref{app:jarz}.  The parameter $\mu_0$ is chosen such that 
\beq
(L/\sigS)^3 \ee^{\beta\mu_0} = \overline{n}_0
\eeq
with $\overline{n}_0$ of order unity being the average number of small particles that would be present, if there were no large particles.
  From grand-canonical simulations at this chemical potential (with fixed large particle configuration $\CC$), we estimate $P_\CC(n=0)$ as the probability to have no small particles at all, from which we compute $\Xi_{\rm S}(\mu_0,\CC)=1/P_\CC(n=0)$.  The chemical potential is increased in steps with $\beta\mu_k=\beta\mu_0+\log (k+1)$,   
  so that $(L/\sigS)^3 \ee^{\beta\mu_k} = (k+1)\overline{n}_0$.  That is, the average number of particles in the system increases linearly with the number of steps.  Between  each step in $\mu$, we perform 1 MCS for the small particles.  We take $\overline{n}_0=(0.25, 0.48, 0.45)$ for $q=(\frac25,\frac14,\frac{2}{13})$. The number of steps $K$ is fixed by the condition that $\mu_K=\muS$, which is the chemical potential for the state point of interest.  
  
The nature of the AO system means that accurate computation of the Jarzynski integral can be simplified.  The system is divided into cells of size $\sigS$.  Before starting the computation, we identify cells that do not overlap at all with any large particle, and also cells where the large particles block the insertion of any small particle.  We do not attempt to insert small particles in any of these cells.  The remaining cells are called \emph{interaction cells} and the number of them is $n_{\rm int}$.  Hence the grand-canonical acceptance probabilities for particle 
insertion $P_{\rm in}$ and removal $P_{\rm del}$ are 
\begin{align}
 P_{\rm in} & = \mathrm{min}\left( 1, \frac{n_{\rm int}v_{\rm cell}}{(n'+1)\sigS^3}\exp(\beta \muS) \right) \, ,
 \\
 P_{\rm del} & = \mathrm{min}\left( 1, \frac{n'\sigS^3}{n_{\rm int}v_{\rm cell}}\exp(-\beta \muS) \right) \, ,
\end{align}
where $n'$ is the number of small particles in the system.
The time taken for the Jarzynski integration depends strongly on the number of small particles that have to be inserted, but a single calculation might take a few minutes, in a typical case.

We compute the Jarzysnki integral based on (\ref{equ:def-I-step}) with this MC method, which yields an unbiased estimate $\ee^{\beta {\cal I}(\CC) }$ of
$\exp[(\ee^{\beta \muS} - \ee^{\beta \mu_0}) {\cal V}^{\rm int}_{\rm a}(C)/\sigS^3]$.  Here, ${\cal V}^{\rm int}_{\rm a}$ is the volume accessible to the small particles, \emph{within the interaction cells}.  The total accessible volume is then
${\cal V}_{\rm a} = n_{\rm free} v_{\rm cell} + {\cal V}^{\rm int}_{\rm a}$, where $n_{\rm free}$ is the number of cells that do not overlap at all with any large particle.  This number is known and so we can replace (\ref{equ:Wm}) with
\begin{multline}
\hat{W}(\CC) = \Xi_{\rm S}(\mu_0,\CC) \ee^{ \beta U_{\rm c}(\CC) }
 \frac{1}{m} \sum_{p=1}^m \ee^{\beta {\cal I}_p(\CC) }
\\ \times \exp\left[  (\ee^{\beta \muS} - \ee^{\beta \mu_0}) n_{\rm free} v_{\rm cell} /\sigS^3  \right]
\end{multline}
which still satisfies (\ref{equ:jarz-result}).

\section{Coarse model with three-body interactions}
\label{app:coarse-three}

\begin{figure*}
\includegraphics[width=145mm]{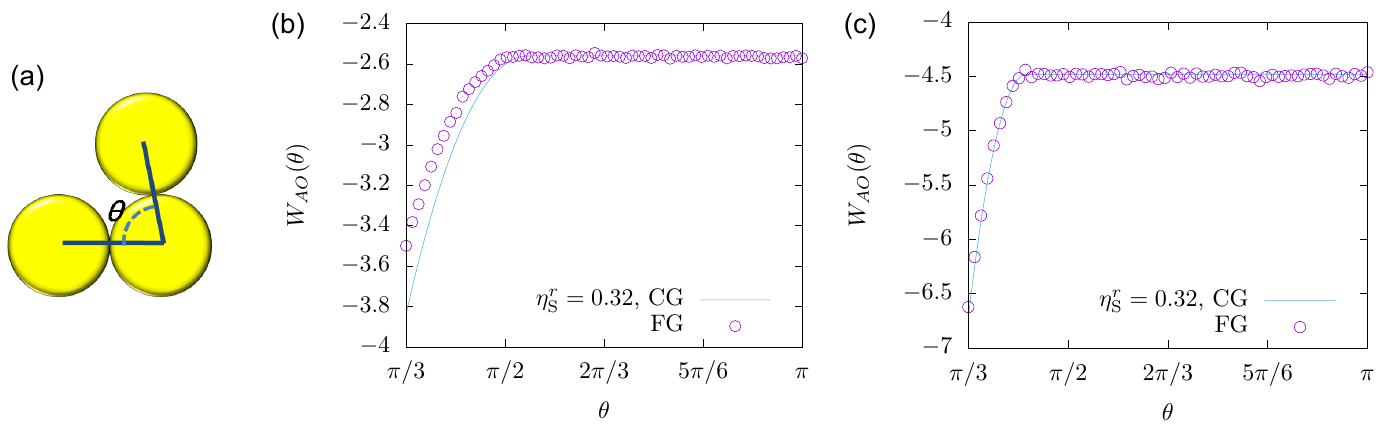}
\caption{Three body interaction in the Asakura-Oosawa system as a function of $\theta$ for $\etaS=0.32$. (a) Schematic illustration of the large-particle geometry used in these calculations.  (b,c) The effective interaction in this three-particle system is shown with symbols (FG) in (a), calculated by Jarzynski integration.
The energy obtained from the pairwise additive two-body interaction potential of Eq.~(\ref{equ:Ucoarse}) is shown with a solid line (CG). In (b) we take $q=\frac12$ and there are significant differences between FG effective interaction and the two-body coarse model.  In (c) then $q=\frac14$ and the differences are smaller.
}
\label{fig:three-body_potential}
\end{figure*}

This Appendix gives extra information on three-body interactions in the system discussed in Sec.~\ref{sec:q25}, which has $q=\frac25$.
As a starting point, Fig.~\ref{fig:ratio25-bad} shows the performance of the TL method, using the coarse model of (\ref{equ:Ucoarse}), which has only
two-body interactions.  In contrast to Fig.~\ref{fig:ratio25}, it is clear that the TL method is not able to correct the coarse-graining errors in this coarse model.  The failure of the method appears as a small number of very large weights $\hat{w}_i$ in Fig.~\ref{fig:ratio25-bad}c, up to $10^{10}$.  The computational effort of  Fig.~\ref{fig:ratio25-bad} is comparable with that of Fig.~\ref{fig:ratio25}, and $\NNf=128,000$ in both cases.  In principle the TL method can be used with any coarse model, but it is clear from (\ref{equ:hat-Delta}) that $\NNf$ should be significantly larger than the largest weights appearing in the calculation, otherwise large errors are expected.   Hence one sees that the TL method is not feasible for the case shown in Fig.~\ref{fig:ratio25-bad}, since one would need $\NNf\gtrsim 10^{10}$.

As discussed in Sec.~\ref{sec:q25}, we analyse the system at $q=(2/5)$ using a CG model with (2+3)-body interactions.
This interaction is represented as in (\ref{equ:U3}), which is a scalar function $U_3$ of three  scalar variables (which are interparticle distances).  The function is symmetric in all its arguments and is non-zero only if all arguments are between $\sigB$ and $\sigB+\sigS$.  In practice we make the slightly stronger restriction that $U_3(x,y,z)=0$ except if $\sigB^2 < (x^2+y^2+z^2) < (\sigB+\sigS)^2$.  We tabulate this function by estimating $U_3(x_i,x_j,x_k)$ with $x_j=j\Delta\sigma$ and $\Delta\sigma=\sigS/10$, and then we estimate the function at generic arguments using linear interpolation.  

The tabulated values of the function are evaluated by Jarzynski integration as in (\ref{equ:jarz-result}), for systems with $N=3$ large particles.  We take $L=3.6\sigB$ and we generate $\NNf=512,000$ configurations of the large particles by MC simulation of the  coarse model (\ref{equ:Ucoarse}).  We perform Jarzynski integration on each of these configurations $\CC$ which yields an estimate of $V_3(\CC) = {6 \etaS} {\cal V}_{\rm a}(\CC)/(\pi\sigS^3\beta) + U_{\rm c}(\CC)$ which is the three-body interaction in (\ref{equ:pm-pc}).  
For each $\CC$ we compute the distances $R_{12},R_{13},R_{23}$ which are the arguments of $U_3$. Then we define $U_3(x_i,x_j,x_k)$ as an average the estimated $V_3(\CC)$ over configurations $\CC$ for which the 3-vector $(R_{12},R_{13},R_{23})$ is inside a cubic box of side $\Delta\sigma$, centred at $(x_i,x_j,x_k)$.  

An alternative approach would be to compute $U_3$ based on a deterministically chosen set of large-particle configurations.  This might reduce the computational effort but it requires prior knowledge of the structure of the three-body potential.  Instead, we  computed the function at a set of representative points and we used these results to build the table.  This ``human-learning'' method is sufficient for this simple three-body potential; machine-learning strategies would also be possible~\cite{Behler2007,Bartok2010,Cheng2019}.

To illustrate the structure of the three-body potential, Fig.~\ref{fig:three-body_potential} shows the energy of the (2+3)-body coarse model for three particles in contact, as a function of the bond angle $\theta$. This calculation is performed by explicitly constructing large-particle configurations as in Fig.~\ref{fig:three-body_potential}(c), and computing their weight factors using (\ref{equ:jarz-result}).  Then $W_{\rm AO}(\CC) = \frac{6 \etaS}{\pi\sigS^3}  [ {\cal V}_{\rm a}(\CC) - {\cal V}_{\rm a}(\CC_0)]$ where $\CC_0$ is a configuration where all particles in the system are well-separated.  This is the exact coarse energy of the (2+3)-body system, which is to be
is compared with the energy of the two-body coarse model.  Small $\theta$ (close to $\pi/3$) corresponds to the case where  all three particles are close to each other.  For $q=(1/2)$, the three-body interaction acts to increase the energy of such configurations, so the effective interaction between large particles is weaker than the prediction of the two-body coarse-model (\ref{equ:Ucoarse}).
For $q=(1/4)$ this effect is weaker, although the result of Fig.~\ref{fig:ratio4}b shows that even these small changes in the interaction potential have significant effects close to criticality.

\section{The bias of the estimator $\hat A$}
\label{app:eps}
We show under the assumptions of Sec.~\ref{sec:con-overview} that the expectation of $\hat{\mathcal E}$ appearing in (\ref{equ:Delta-E}) is of order $\NNf^{-1}$.
By definition, $\hat{\mathcal{E}}$ is an importance sampling estimator which converges to $0$ almost surely for $\NNf \to \infty$.
In (\ref{equ:w-eps}) we have seen that the normalisation of $\hat{w}$ slightly perturbs the weights appearing in $\hat{\mathcal{E}}$ , i.e.\ in general $\eps \neq 0$.
This results in a bias which we can estimate by following  Ref.~\onlinecite{Agapiou2017}.  For completeness, we reproduce here the steps used in the proof of their Theorem 2.1, adapted to our notation.

We introduce the short-hand notation
\beq
\delta A(\CC_j) = A(\CC_j) - \langle A \rangle_{\rm f} \; .
\eeq
Also note  
from (\ref{equ:jarz-result}) that
\beq
\langle {\hat{W}(\CC_i)}/{\xi} \rangle_{\rm J} = \frac{p_{\rm m}(\CC_i)}{ p_{\rm c}(\CC_i) } .
\label{equ:Wp}
\eeq
Using the definition of $\hat{\cal E}$ with (\ref{equ:Amc},\ref{equ:w-eps},\ref{equ:Wp}) yields
\beq
\langle \hat{\mathcal{E}} \rangle = \left\langle \frac{-\eps}{1+\eps}\frac{1}{\NNf} \sum_j \delta A(\CC_j) 
\hat{W}(\CC_j)/\xi     \right\rangle
\eeq
We decompose the expectation value into contributions from $\eps\leq-\frac12$ and $\eps>-\frac12$, so that
\beq
    |\langle \hat{\mathcal E} \rangle | \leq | \langle \hat{\mathcal E} 1_{\{2(\eps+1) > 1\}} \rangle |
            + | \langle \hat{\mathcal E} 1_{\{2 (\eps+1) \leq 1\}} \rangle | \, . \label{equ:E-bound}
\eeq
where the indicator function $1_{\{X\}}$ is equal to unity if $X$ is true, and zero otherwise.
For the first term on the right hand side of (\ref{equ:E-bound}), we use that $|\eps/(1+\eps)|\leq 2|\eps|$ for $\eps>-\frac12$ to obtain
\begin{equation}
| \langle \hat{\mathcal E} 1_{\{2(\eps+1) > 1\}} \rangle | 
   \leq 2 \Big \langle |\eps| \cdot \Big| \frac{1}{\NNf} \sum_j \delta A(\CC_j) \hat{W}(\CC_j)/\xi \Big| \Big \rangle 
   \eeq
The right hand side is less than or equal to
\beq
2 \langle \eps^2 \rangle^{1/2} \cdot \Big\langle \Big( \frac{1}{\NNf} \sum_j \delta A(\CC_j) \hat{W}(\CC_j)/\xi \Big)^2 \Big\rangle^{1/2}
\eeq
which follows from the  Cauchy-Schwarz inequality.
Using this fact together with (\ref{equ:def-eps}) and that the $\CC_i$ are all independent, we find
   \begin{multline}
   | \langle \hat{\mathcal E} 1_{\{2(\eps+1) > 1\}} \rangle |
    \\ \leq \frac{2}{\NNf\xi^2} 
    \left( 
    \langle \hat{W}(\CC_j)^2 \rangle
    \cdot 
    \langle [\delta A(\CC_j) \hat{W}(\CC_j)]^2 \rangle \right)^{1/2} 
\label{equ:bias-intermediate}
\end{multline}
Then (\ref{equ:bounded-A}) yields
\beq
| \langle \hat{\mathcal E} 1_{\{2(\eps+1) > 1\}} \rangle | \leq \frac{2a}{\NNf\xi^2}\langle\hat{W}(\CC_j)^2\rangle \; .
\label{equ:E-part1}
\eeq

For the second term on the right hand side of (\ref{equ:E-bound}), one uses (\ref{equ:ratio},\ref{equ:bounded-A},\ref{equ:E-hat-def}) to show that $|\hat{\mathcal E}| \leq a$.   Hence
\beq
| \langle \hat{\mathcal E} 1_{\{2 (\eps+1) \leq 1\}} \rangle |
    \leq a \,\mathrm{Prob}(\eps \leq -1/2).
\eeq
This probability can be bounded by Chebyshev's inequality which implies that for any mean-zero random variable $X$ and any number $x>0$ then $\mathrm{Prob}(X\leq -x) \leq \mathrm{Var}(X)/x^2$.  Hence
\beq
   | \langle \hat{\mathcal E} 1_{\{2 (\eps+1) \leq 1\}} \rangle | \leq 4 a \mathrm{Var}(\eps) = \frac{4a}{\xi^2 \NNf} \mathrm{Var}(\hat{W}(\CC_j)).
    \label{equ:E-part2}
\eeq
Combining (\ref{equ:E-bound},\ref{equ:E-part1},\ref{equ:E-part2}) we have
\beq
|\langle \hat{\mathcal{E}} \rangle| \leq \frac{6a}{\xi^2 \NNf} 
\langle\hat{W}(\CC_j)^2\rangle
\label{equ:bias-bound-final}
\eeq
which is an explicit bound that holds for all $\NNf$. 
The right hand side is finite by (\ref{equ:OmTot}) so
$
\langle \hat{\mathcal{E}}\rangle %
$
is
$O(\NNf^{-1})$
as advertised above.  With (\ref{equ:Delta-E}) this establishes (\ref{equ:mean-Delta}).

We note in passing that if the weights have zero variance then $\hat{\cal E}=0$ strictly and there is no bias, although the bound (\ref{equ:bias-bound-final}) does not capture this fact.  In fact this bound may be improved by refining the estimate of $\langle\eps^2\rangle$ used in (\ref{equ:bias-intermediate}).

\end{appendix}

\bibliography{multi}

\begin{thebibliography}{62}%
\makeatletter
\providecommand \@ifxundefined [1]{%
 \@ifx{#1\undefined}
}%
\providecommand \@ifnum [1]{%
 \ifnum #1\expandafter \@firstoftwo
 \else \expandafter \@secondoftwo
 \fi
}%
\providecommand \@ifx [1]{%
 \ifx #1\expandafter \@firstoftwo
 \else \expandafter \@secondoftwo
 \fi
}%
\providecommand \natexlab [1]{#1}%
\providecommand \enquote  [1]{``#1''}%
\providecommand \bibnamefont  [1]{#1}%
\providecommand \bibfnamefont [1]{#1}%
\providecommand \citenamefont [1]{#1}%
\providecommand \href@noop [0]{\@secondoftwo}%
\providecommand \href [0]{\begingroup \@sanitize@url \@href}%
\providecommand \@href[1]{\@@startlink{#1}\@@href}%
\providecommand \@@href[1]{\endgroup#1\@@endlink}%
\providecommand \@sanitize@url [0]{\catcode `\\12\catcode `\$12\catcode
  `\&12\catcode `\#12\catcode `\^12\catcode `\_12\catcode `\%12\relax}%
\providecommand \@@startlink[1]{}%
\providecommand \@@endlink[0]{}%
\providecommand \url  [0]{\begingroup\@sanitize@url \@url }%
\providecommand \@url [1]{\endgroup\@href {#1}{\urlprefix }}%
\providecommand \urlprefix  [0]{URL }%
\providecommand \Eprint [0]{\href }%
\providecommand \doibase [0]{http://dx.doi.org/}%
\providecommand \selectlanguage [0]{\@gobble}%
\providecommand \bibinfo  [0]{\@secondoftwo}%
\providecommand \bibfield  [0]{\@secondoftwo}%
\providecommand \translation [1]{[#1]}%
\providecommand \BibitemOpen [0]{}%
\providecommand \bibitemStop [0]{}%
\providecommand \bibitemNoStop [0]{.\EOS\space}%
\providecommand \EOS [0]{\spacefactor3000\relax}%
\providecommand \BibitemShut  [1]{\csname bibitem#1\endcsname}%
\let\auto@bib@innerbib\@empty
\bibitem [{\citenamefont {Likos}(2001)}]{Likos2001}%
  \BibitemOpen
  \bibfield  {author} {\bibinfo {author} {\bibfnamefont {C.~N.}\ \bibnamefont
  {Likos}},\ }\bibfield  {title} {\enquote {\bibinfo {title} {Effective
  interactions in soft condensed matter physics},}\ }\href {\doibase
  10.1016/S0370-1573(00)00141-1} {\bibfield  {journal} {\bibinfo  {journal}
  {Phys. Rep.}\ }\textbf {\bibinfo {volume} {348}},\ \bibinfo {pages} {267 }
  (\bibinfo {year} {2001})}\BibitemShut {NoStop}%
\bibitem [{\citenamefont {Pak}\ and\ \citenamefont {Voth}(2018)}]{Pak2018}%
  \BibitemOpen
  \bibfield  {author} {\bibinfo {author} {\bibfnamefont {A.~J.}\ \bibnamefont
  {Pak}}\ and\ \bibinfo {author} {\bibfnamefont {G.~A.}\ \bibnamefont {Voth}},\
  }\bibfield  {title} {\enquote {\bibinfo {title} {Advances in coarse-grained
  modeling of macromolecular complexes},}\ }\href
  {https://doi.org/10.1016/j.sbi.2018.11.005} {\bibfield  {journal} {\bibinfo
  {journal} {Curr. Opin. Struct. Biol.}\ }\textbf {\bibinfo {volume} {52}},\
  \bibinfo {pages} {119} (\bibinfo {year} {2018})}\BibitemShut {NoStop}%
\bibitem [{\citenamefont {Hagan}\ and\ \citenamefont
  {Chandler}(2006)}]{Hagan2006}%
  \BibitemOpen
  \bibfield  {author} {\bibinfo {author} {\bibfnamefont {M.}~\bibnamefont
  {Hagan}}\ and\ \bibinfo {author} {\bibfnamefont {D.}~\bibnamefont
  {Chandler}},\ }\bibfield  {title} {\enquote {\bibinfo {title} {Dynamic
  pathways for viral capsid assembly},}\ }\href {\doibase
  10.1529/biophysj.105.076851} {\bibfield  {journal} {\bibinfo  {journal}
  {Biophys. J.}\ }\textbf {\bibinfo {volume} {91}},\ \bibinfo {pages} {42}
  (\bibinfo {year} {2006})}\BibitemShut {NoStop}%
\bibitem [{\citenamefont {Haxton}\ and\ \citenamefont
  {Whitelam}(2012)}]{Haxton2012}%
  \BibitemOpen
  \bibfield  {author} {\bibinfo {author} {\bibfnamefont {T.~K.}\ \bibnamefont
  {Haxton}}\ and\ \bibinfo {author} {\bibfnamefont {S.}~\bibnamefont
  {Whitelam}},\ }\bibfield  {title} {\enquote {\bibinfo {title} {Design rules
  for the self-assembly of a protein crystal},}\ }\href {\doibase
  10.1039/C2SM07436B} {\bibfield  {journal} {\bibinfo  {journal} {Soft Matter}\
  }\textbf {\bibinfo {volume} {8}},\ \bibinfo {pages} {3558} (\bibinfo {year}
  {2012})}\BibitemShut {NoStop}%
\bibitem [{\citenamefont {Saric}\ \emph {et~al.}(2014)\citenamefont {Saric},
  \citenamefont {Chebaro}, \citenamefont {Knowles},\ and\ \citenamefont
  {Frenkel}}]{Saric2014}%
  \BibitemOpen
  \bibfield  {author} {\bibinfo {author} {\bibfnamefont {A.}~\bibnamefont
  {Saric}}, \bibinfo {author} {\bibfnamefont {Y.~C.}\ \bibnamefont {Chebaro}},
  \bibinfo {author} {\bibfnamefont {T.~P.~J.}\ \bibnamefont {Knowles}}, \ and\
  \bibinfo {author} {\bibfnamefont {D.}~\bibnamefont {Frenkel}},\ }\bibfield
  {title} {\enquote {\bibinfo {title} {Crucial role of nonspecific interactions
  in amyloid nucleation},}\ }\href {\doibase 10.1073/pnas.1410159111}
  {\bibfield  {journal} {\bibinfo  {journal} {Proc. Natl. Acad. Sci. USA}\
  }\textbf {\bibinfo {volume} {111}},\ \bibinfo {pages} {17869} (\bibinfo
  {year} {2014})}\BibitemShut {NoStop}%
\bibitem [{\citenamefont {Noid}\ \emph {et~al.}(2008)\citenamefont {Noid},
  \citenamefont {Chu}, \citenamefont {Ayton}, \citenamefont {Krishna},
  \citenamefont {Izvekov}, \citenamefont {Voth}, \citenamefont {Das},\ and\
  \citenamefont {Andersen}}]{Noid2008}%
  \BibitemOpen
  \bibfield  {author} {\bibinfo {author} {\bibfnamefont {W.~G.}\ \bibnamefont
  {Noid}}, \bibinfo {author} {\bibfnamefont {J.-W.}\ \bibnamefont {Chu}},
  \bibinfo {author} {\bibfnamefont {G.~S.}\ \bibnamefont {Ayton}}, \bibinfo
  {author} {\bibfnamefont {V.}~\bibnamefont {Krishna}}, \bibinfo {author}
  {\bibfnamefont {S.}~\bibnamefont {Izvekov}}, \bibinfo {author} {\bibfnamefont
  {G.~A.}\ \bibnamefont {Voth}}, \bibinfo {author} {\bibfnamefont
  {A.}~\bibnamefont {Das}}, \ and\ \bibinfo {author} {\bibfnamefont {H.~C.}\
  \bibnamefont {Andersen}},\ }\bibfield  {title} {\enquote {\bibinfo {title}
  {The multiscale coarse-graining method. {I.} {A} rigorous bridge between
  atomistic and coarse-grained models},}\ }\href {\doibase 10.1063/1.2938860}
  {\bibfield  {journal} {\bibinfo  {journal} {J. Chem. Phys.}\ }\textbf
  {\bibinfo {volume} {128}},\ \bibinfo {pages} {244114} (\bibinfo {year}
  {2008})}\BibitemShut {NoStop}%
\bibitem [{\citenamefont {Praprotnik}, \citenamefont {Delle~Site},\ and\
  \citenamefont {Kremer}(2007)}]{Praprotnik2007}%
  \BibitemOpen
  \bibfield  {author} {\bibinfo {author} {\bibfnamefont {M.}~\bibnamefont
  {Praprotnik}}, \bibinfo {author} {\bibfnamefont {L.}~\bibnamefont
  {Delle~Site}}, \ and\ \bibinfo {author} {\bibfnamefont {K.}~\bibnamefont
  {Kremer}},\ }\bibfield  {title} {\enquote {\bibinfo {title} {A macromolecule
  in a solvent: Adaptive resolution molecular dynamics simulation},}\ }\href
  {http://dx.doi.org/10.1063/1.2714540} {\bibfield  {journal} {\bibinfo
  {journal} {J. Chem. Phys.}\ }\textbf {\bibinfo {volume} {126}},\ \bibinfo
  {pages} {134902} (\bibinfo {year} {2007})}\BibitemShut {NoStop}%
\bibitem [{\citenamefont {Ouldridge}, \citenamefont {Louis},\ and\
  \citenamefont {Doye}(2011)}]{Ouldridge2011}%
  \BibitemOpen
  \bibfield  {author} {\bibinfo {author} {\bibfnamefont {T.~E.}\ \bibnamefont
  {Ouldridge}}, \bibinfo {author} {\bibfnamefont {A.~A.}\ \bibnamefont
  {Louis}}, \ and\ \bibinfo {author} {\bibfnamefont {J.~P.~K.}\ \bibnamefont
  {Doye}},\ }\bibfield  {title} {\enquote {\bibinfo {title} {Structural,
  mechanical, and thermodynamic properties of a coarse-grained {DNA} model},}\
  }\href {\doibase 10.1063/1.3552946} {\bibfield  {journal} {\bibinfo
  {journal} {J. Chem. Phys.}\ }\textbf {\bibinfo {volume} {134}},\ \bibinfo
  {pages} {085101} (\bibinfo {year} {2011})}\BibitemShut {NoStop}%
\bibitem [{\citenamefont {Mladek}\ \emph {et~al.}(2013)\citenamefont {Mladek},
  \citenamefont {Fornleitner}, \citenamefont {Martinez-Veracoechea},
  \citenamefont {Dawid},\ and\ \citenamefont {Frenkel}}]{Mladek2013}%
  \BibitemOpen
  \bibfield  {author} {\bibinfo {author} {\bibfnamefont {B.~M.}\ \bibnamefont
  {Mladek}}, \bibinfo {author} {\bibfnamefont {J.}~\bibnamefont {Fornleitner}},
  \bibinfo {author} {\bibfnamefont {F.~J.}\ \bibnamefont
  {Martinez-Veracoechea}}, \bibinfo {author} {\bibfnamefont {A.}~\bibnamefont
  {Dawid}}, \ and\ \bibinfo {author} {\bibfnamefont {D.}~\bibnamefont
  {Frenkel}},\ }\bibfield  {title} {\enquote {\bibinfo {title} {Procedure to
  construct a multi-scale coarse-grained model of {DNA}-coated colloids from
  experimental data},}\ }\href {\doibase 10.1039/C3SM50701G} {\bibfield
  {journal} {\bibinfo  {journal} {Soft Matter}\ }\textbf {\bibinfo {volume}
  {9}},\ \bibinfo {pages} {7342} (\bibinfo {year} {2013})}\BibitemShut
  {NoStop}%
\bibitem [{\citenamefont {Karplus}(2014)}]{Karplus2014}%
  \BibitemOpen
  \bibfield  {author} {\bibinfo {author} {\bibfnamefont {M.}~\bibnamefont
  {Karplus}},\ }\bibfield  {title} {\enquote {\bibinfo {title} {Development of
  multiscale models for complex chemical systems: From {H}+{H$_2$} to
  biomolecules ({N}obel lecture)},}\ }\href {\doibase 10.1002/anie.201403924}
  {\bibfield  {journal} {\bibinfo  {journal} {Angew. Chem.: Intl. Ed.}\
  }\textbf {\bibinfo {volume} {53}},\ \bibinfo {pages} {9992} (\bibinfo {year}
  {2014})}\BibitemShut {NoStop}%
\bibitem [{\citenamefont {Warshel}(2014)}]{Warshel2014}%
  \BibitemOpen
  \bibfield  {author} {\bibinfo {author} {\bibfnamefont {A.}~\bibnamefont
  {Warshel}},\ }\bibfield  {title} {\enquote {\bibinfo {title} {Multiscale
  modeling of biological functions: From enzymes to molecular machines ({N}obel
  lecture)},}\ }\href {\doibase 10.1002/anie.201403689} {\bibfield  {journal}
  {\bibinfo  {journal} {Angew. Chem.: Intl. Ed.}\ }\textbf {\bibinfo {volume}
  {53}},\ \bibinfo {pages} {10020} (\bibinfo {year} {2014})}\BibitemShut
  {NoStop}%
\bibitem [{\citenamefont {John}\ and\ \citenamefont
  {Cs{\'a}nyi}(2017)}]{John2017}%
  \BibitemOpen
  \bibfield  {author} {\bibinfo {author} {\bibfnamefont {S.~T.}\ \bibnamefont
  {John}}\ and\ \bibinfo {author} {\bibfnamefont {G.}~\bibnamefont
  {Cs{\'a}nyi}},\ }\bibfield  {title} {\enquote {\bibinfo {title} {Many-body
  coarse-grained interactions using {G}aussian approximation potentials},}\
  }\href {\doibase 10.1021/acs.jpcb.7b09636} {\bibfield  {journal} {\bibinfo
  {journal} {J. Phys. Chem. B}\ }\textbf {\bibinfo {volume} {121}},\ \bibinfo
  {pages} {10934} (\bibinfo {year} {2017})}\BibitemShut {NoStop}%
\bibitem [{\citenamefont {Behler}\ and\ \citenamefont
  {Parrinello}(2007)}]{Behler2007}%
  \BibitemOpen
  \bibfield  {author} {\bibinfo {author} {\bibfnamefont {J.}~\bibnamefont
  {Behler}}\ and\ \bibinfo {author} {\bibfnamefont {M.}~\bibnamefont
  {Parrinello}},\ }\bibfield  {title} {\enquote {\bibinfo {title} {Generalized
  neural-network representation of high-dimensional potential-energy
  surfaces},}\ }\href {\doibase 10.1103/PhysRevLett.98.146401} {\bibfield
  {journal} {\bibinfo  {journal} {Phys. Rev. Lett.}\ }\textbf {\bibinfo
  {volume} {98}},\ \bibinfo {pages} {146401} (\bibinfo {year}
  {2007})}\BibitemShut {NoStop}%
\bibitem [{\citenamefont {Bart{\'o}k}\ \emph {et~al.}(2010)\citenamefont
  {Bart{\'o}k}, \citenamefont {Payne}, \citenamefont {Kondor},\ and\
  \citenamefont {Cs{\'a}nyi}}]{Bartok2010}%
  \BibitemOpen
  \bibfield  {author} {\bibinfo {author} {\bibfnamefont {A.~P.}\ \bibnamefont
  {Bart{\'o}k}}, \bibinfo {author} {\bibfnamefont {M.~C.}\ \bibnamefont
  {Payne}}, \bibinfo {author} {\bibfnamefont {R.}~\bibnamefont {Kondor}}, \
  and\ \bibinfo {author} {\bibfnamefont {G.}~\bibnamefont {Cs{\'a}nyi}},\
  }\bibfield  {title} {\enquote {\bibinfo {title} {Gaussian approximation
  potentials: The accuracy of quantum mechanics, without the electrons},}\
  }\href {\doibase 10.1103/PhysRevLett.104.136403} {\bibfield  {journal}
  {\bibinfo  {journal} {Phys. Rev. Lett.}\ }\textbf {\bibinfo {volume} {104}},\
  \bibinfo {pages} {136403} (\bibinfo {year} {2010})}\BibitemShut {NoStop}%
\bibitem [{\citenamefont {Cheng}\ \emph {et~al.}(2019)\citenamefont {Cheng},
  \citenamefont {Engel}, \citenamefont {Behler}, \citenamefont {Dellago},\ and\
  \citenamefont {Ceriotti}}]{Cheng2019}%
  \BibitemOpen
  \bibfield  {author} {\bibinfo {author} {\bibfnamefont {B.}~\bibnamefont
  {Cheng}}, \bibinfo {author} {\bibfnamefont {E.~A.}\ \bibnamefont {Engel}},
  \bibinfo {author} {\bibfnamefont {J.}~\bibnamefont {Behler}}, \bibinfo
  {author} {\bibfnamefont {C.}~\bibnamefont {Dellago}}, \ and\ \bibinfo
  {author} {\bibfnamefont {M.}~\bibnamefont {Ceriotti}},\ }\bibfield  {title}
  {\enquote {\bibinfo {title} {Ab initio thermodynamics of liquid and solid
  water},}\ }\href@noop {} {\bibfield  {journal} {\bibinfo  {journal} {Proc.
  Natl. Acad. Sci. USA}\ }\textbf {\bibinfo {volume} {116}},\ \bibinfo {pages}
  {1110} (\bibinfo {year} {2019})}\BibitemShut {NoStop}%
\bibitem [{\citenamefont {Heinrich}(1998)}]{heinrich1998monte}%
  \BibitemOpen
  \bibfield  {author} {\bibinfo {author} {\bibfnamefont {S.}~\bibnamefont
  {Heinrich}},\ }\bibfield  {title} {\enquote {\bibinfo {title} {{M}onte
  {C}arlo complexity of global solution of integral equations},}\ }\href
  {\doibase 10.1006/jcom.1998.0471} {\bibfield  {journal} {\bibinfo  {journal}
  {J. Complex.}\ }\textbf {\bibinfo {volume} {14}},\ \bibinfo {pages} {151}
  (\bibinfo {year} {1998})}\BibitemShut {NoStop}%
\bibitem [{\citenamefont {Giles}(2008)}]{Giles2008}%
  \BibitemOpen
  \bibfield  {author} {\bibinfo {author} {\bibfnamefont {M.~B.}\ \bibnamefont
  {Giles}},\ }\bibfield  {title} {\enquote {\bibinfo {title} {Multi-level
  {M}onte {C}arlo path simulation},}\ }\href {\doibase 10.1287/opre.1070.0496}
  {\bibfield  {journal} {\bibinfo  {journal} {Operations Res.}\ }\textbf
  {\bibinfo {volume} {56}},\ \bibinfo {pages} {607} (\bibinfo {year}
  {2008})}\BibitemShut {NoStop}%
\bibitem [{\citenamefont {Anderson}\ and\ \citenamefont
  {Higham}(2012)}]{Anderson2012}%
  \BibitemOpen
  \bibfield  {author} {\bibinfo {author} {\bibfnamefont {D.~F.}\ \bibnamefont
  {Anderson}}\ and\ \bibinfo {author} {\bibfnamefont {D.~J.}\ \bibnamefont
  {Higham}},\ }\bibfield  {title} {\enquote {\bibinfo {title} {Multilevel
  {M}onte {C}arlo for continuous time {M}arkov chains, with applications in
  biochemical kinetics},}\ }\href@noop {} {\bibfield  {journal} {\bibinfo
  {journal} {SIAM Multiscale Modelling and Simulation}\ }\textbf {\bibinfo
  {volume} {10}},\ \bibinfo {pages} {146} (\bibinfo {year} {2012})}\BibitemShut
  {NoStop}%
\bibitem [{\citenamefont {Hoang}, \citenamefont {Schwab},\ and\ \citenamefont
  {Stuart}(2013)}]{Hoang2013}%
  \BibitemOpen
  \bibfield  {author} {\bibinfo {author} {\bibfnamefont {V.~H.}\ \bibnamefont
  {Hoang}}, \bibinfo {author} {\bibfnamefont {C.}~\bibnamefont {Schwab}}, \
  and\ \bibinfo {author} {\bibfnamefont {A.~M.}\ \bibnamefont {Stuart}},\
  }\bibfield  {title} {\enquote {\bibinfo {title} {Complexity analysis of
  accelerated {MCMC} methods for {B}ayesian inversion},}\ }\href {\doibase
  doi:10.1088/0266-5611/29/8/085010} {\bibfield  {journal} {\bibinfo  {journal}
  {Inverse Problems}\ }\textbf {\bibinfo {volume} {29}},\ \bibinfo {pages}
  {85010} (\bibinfo {year} {2013})}\BibitemShut {NoStop}%
\bibitem [{\citenamefont {Dodwell}\ \emph {et~al.}(2015)\citenamefont
  {Dodwell}, \citenamefont {Ketelsen}, \citenamefont {Scheichl},\ and\
  \citenamefont {Teckentrup}}]{Dodwell2015}%
  \BibitemOpen
  \bibfield  {author} {\bibinfo {author} {\bibfnamefont {T.~J.}\ \bibnamefont
  {Dodwell}}, \bibinfo {author} {\bibfnamefont {C.}~\bibnamefont {Ketelsen}},
  \bibinfo {author} {\bibfnamefont {R.}~\bibnamefont {Scheichl}}, \ and\
  \bibinfo {author} {\bibfnamefont {A.~L.}\ \bibnamefont {Teckentrup}},\
  }\bibfield  {title} {\enquote {\bibinfo {title} {A hierarchical multilevel
  {M}arkov chain {M}onte {C}arlo algorithm with applications to uncertainty
  quantification in subsurface flow},}\ }\href {\doibase 10.1137/130915005}
  {\bibfield  {journal} {\bibinfo  {journal} {SIAM/ASA J. Uncertainty Quant.}\
  }\textbf {\bibinfo {volume} {3}},\ \bibinfo {pages} {1075} (\bibinfo {year}
  {2015})}\BibitemShut {NoStop}%
\bibitem [{\citenamefont {Beskos}\ \emph {et~al.}(2017)\citenamefont {Beskos},
  \citenamefont {Jasra}, \citenamefont {Law}, \citenamefont {Tempone},\ and\
  \citenamefont {Zhou}}]{beskos2017multilevel}%
  \BibitemOpen
  \bibfield  {author} {\bibinfo {author} {\bibfnamefont {A.}~\bibnamefont
  {Beskos}}, \bibinfo {author} {\bibfnamefont {A.}~\bibnamefont {Jasra}},
  \bibinfo {author} {\bibfnamefont {K.}~\bibnamefont {Law}}, \bibinfo {author}
  {\bibfnamefont {R.}~\bibnamefont {Tempone}}, \ and\ \bibinfo {author}
  {\bibfnamefont {Y.}~\bibnamefont {Zhou}},\ }\bibfield  {title} {\enquote
  {\bibinfo {title} {Multilevel sequential {M}onte {C}arlo samplers},}\ }\href
  {\doibase 10.1016/j.spa.2016.08.004} {\bibfield  {journal} {\bibinfo
  {journal} {Stoch. Proc. Appl.}\ }\textbf {\bibinfo {volume} {127}},\ \bibinfo
  {pages} {1417} (\bibinfo {year} {2017})}\BibitemShut {NoStop}%
\bibitem [{\citenamefont {Frenkel}\ and\ \citenamefont
  {Smit}(2001)}]{frenkel-smit}%
  \BibitemOpen
  \bibfield  {author} {\bibinfo {author} {\bibfnamefont {D.}~\bibnamefont
  {Frenkel}}\ and\ \bibinfo {author} {\bibfnamefont {B.}~\bibnamefont {Smit}},\
  }\href@noop {} {\emph {\bibinfo {title} {Understanding Molecular Simulation:
  From Algorithms to Applications}}}\ (\bibinfo  {publisher} {Elsevier},\
  \bibinfo {year} {2001})\BibitemShut {NoStop}%
\bibitem [{\citenamefont {Bruce}\ and\ \citenamefont
  {Wilding}(2003)}]{Bruce2003}%
  \BibitemOpen
  \bibfield  {author} {\bibinfo {author} {\bibfnamefont {A.~D.}\ \bibnamefont
  {Bruce}}\ and\ \bibinfo {author} {\bibfnamefont {N.~B.}\ \bibnamefont
  {Wilding}},\ }\bibfield  {title} {\enquote {\bibinfo {title} {Computational
  strategies for mapping equilibrium phase diagrams},}\ }\href@noop {}
  {\bibfield  {journal} {\bibinfo  {journal} {Adv. Chem. Phys.}\ }\textbf
  {\bibinfo {volume} {127}},\ \bibinfo {pages} {1} (\bibinfo {year}
  {2003})}\BibitemShut {NoStop}%
\bibitem [{\citenamefont {Zwanzig}(1954)}]{Zwanzig1954}%
  \BibitemOpen
  \bibfield  {author} {\bibinfo {author} {\bibfnamefont {R.~W.}\ \bibnamefont
  {Zwanzig}},\ }\bibfield  {title} {\enquote {\bibinfo {title}
  {High‐temperature equation of state by a perturbation method. {I}.
  {N}onpolar gases},}\ }\href {\doibase 10.1063/1.1740409} {\bibfield
  {journal} {\bibinfo  {journal} {J. Chem. Phys.}\ }\textbf {\bibinfo {volume}
  {22}},\ \bibinfo {pages} {1420} (\bibinfo {year} {1954})}\BibitemShut
  {NoStop}%
\bibitem [{\citenamefont {Hummer}\ and\ \citenamefont
  {Szabo}(2001)}]{Hummer2001}%
  \BibitemOpen
  \bibfield  {author} {\bibinfo {author} {\bibfnamefont {G.}~\bibnamefont
  {Hummer}}\ and\ \bibinfo {author} {\bibfnamefont {A.}~\bibnamefont {Szabo}},\
  }\bibfield  {title} {\enquote {\bibinfo {title} {Free energy reconstruction
  from nonequilibrium single-molecule pulling experiments},}\ }\href {\doibase
  10.1073/pnas.071034098} {\bibfield  {journal} {\bibinfo  {journal} {Proc.
  Natl. Acad. Sci. USA}\ }\textbf {\bibinfo {volume} {98}},\ \bibinfo {pages}
  {3658} (\bibinfo {year} {2001})}\BibitemShut {NoStop}%
\bibitem [{\citenamefont {Jarzynski}(1997)}]{Jarzynski1997}%
  \BibitemOpen
  \bibfield  {author} {\bibinfo {author} {\bibfnamefont {C.}~\bibnamefont
  {Jarzynski}},\ }\bibfield  {title} {\enquote {\bibinfo {title}
  {Nonequilibrium equality for free energy differences},}\ }\href@noop {}
  {\bibfield  {journal} {\bibinfo  {journal} {Phys. Rev. Lett.}\ }\textbf
  {\bibinfo {volume} {78}},\ \bibinfo {pages} {2690} (\bibinfo {year}
  {1997})}\BibitemShut {NoStop}%
\bibitem [{\citenamefont {Crooks}(2000)}]{Crooks2000}%
  \BibitemOpen
  \bibfield  {author} {\bibinfo {author} {\bibfnamefont {G.~E.}\ \bibnamefont
  {Crooks}},\ }\bibfield  {title} {\enquote {\bibinfo {title} {Path-ensemble
  averages in systems driven far from equilibrium},}\ }\href@noop {} {\bibfield
   {journal} {\bibinfo  {journal} {Phys. Rev. E}\ }\textbf {\bibinfo {volume}
  {61}},\ \bibinfo {pages} {2361} (\bibinfo {year} {2000})}\BibitemShut
  {NoStop}%
\bibitem [{\citenamefont {Neal}(2001)}]{Neal2001}%
  \BibitemOpen
  \bibfield  {author} {\bibinfo {author} {\bibfnamefont {R.~M.}\ \bibnamefont
  {Neal}},\ }\bibfield  {title} {\enquote {\bibinfo {title} {Annealed
  importance sampling},}\ }\href {\doibase 10.1023/A:1008923215028} {\bibfield
  {journal} {\bibinfo  {journal} {Statistics and Computing}\ }\textbf {\bibinfo
  {volume} {11}},\ \bibinfo {pages} {125} (\bibinfo {year} {2001})}\BibitemShut
  {NoStop}%
\bibitem [{\citenamefont {Asakura}\ and\ \citenamefont
  {Oosawa}(1954)}]{Asakura1954}%
  \BibitemOpen
  \bibfield  {author} {\bibinfo {author} {\bibfnamefont {S.}~\bibnamefont
  {Asakura}}\ and\ \bibinfo {author} {\bibfnamefont {F.}~\bibnamefont
  {Oosawa}},\ }\bibfield  {title} {\enquote {\bibinfo {title} {On interaction
  between two bodies immersed in a solution of macromolecules},}\ }\href
  {\doibase 10.1063/1.1740347} {\bibfield  {journal} {\bibinfo  {journal} {J.
  Chem. Phys.}\ }\textbf {\bibinfo {volume} {22}},\ \bibinfo {pages} {1255}
  (\bibinfo {year} {1954})}\BibitemShut {NoStop}%
\bibitem [{\citenamefont {Binder}, \citenamefont {Virnau},\ and\ \citenamefont
  {Statt}(2014)}]{Binder2014}%
  \BibitemOpen
  \bibfield  {author} {\bibinfo {author} {\bibfnamefont {K.}~\bibnamefont
  {Binder}}, \bibinfo {author} {\bibfnamefont {P.}~\bibnamefont {Virnau}}, \
  and\ \bibinfo {author} {\bibfnamefont {A.}~\bibnamefont {Statt}},\ }\bibfield
   {title} {\enquote {\bibinfo {title} {Perspective: The {A}sakura {O}osawa
  model: A colloid prototype for bulk and interfacial phase behavior},}\
  }\href@noop {} {\bibfield  {journal} {\bibinfo  {journal} {J. Chem. Phys.}\
  }\textbf {\bibinfo {volume} {141}},\ \bibinfo {pages} {140901} (\bibinfo
  {year} {2014})}\BibitemShut {NoStop}%
\bibitem [{\citenamefont {Poon}(2002)}]{Poon2002}%
  \BibitemOpen
  \bibfield  {author} {\bibinfo {author} {\bibfnamefont {W.~C.~K.}\
  \bibnamefont {Poon}},\ }\bibfield  {title} {\enquote {\bibinfo {title} {The
  physics of a model colloid--polymer mixture},}\ }\href@noop {} {\bibfield
  {journal} {\bibinfo  {journal} {J. Phys.: Cond. Matt.}\ }\textbf {\bibinfo
  {volume} {14}},\ \bibinfo {pages} {R859} (\bibinfo {year}
  {2002})}\BibitemShut {NoStop}%
\bibitem [{\citenamefont {Dijkstra}, \citenamefont {Brader},\ and\
  \citenamefont {Evans}(1999)}]{Dijkstra1999-jpcm}%
  \BibitemOpen
  \bibfield  {author} {\bibinfo {author} {\bibfnamefont {M.}~\bibnamefont
  {Dijkstra}}, \bibinfo {author} {\bibfnamefont {J.~M.}\ \bibnamefont
  {Brader}}, \ and\ \bibinfo {author} {\bibfnamefont {R.}~\bibnamefont
  {Evans}},\ }\bibfield  {title} {\enquote {\bibinfo {title} {Phase behaviour
  and structure of model colloid-polymer mixtures},}\ }\href {\doibase
  10.1088/0953-8984/11/50/304} {\bibfield  {journal} {\bibinfo  {journal} {J.
  Phys.: Cond. Matt.}\ }\textbf {\bibinfo {volume} {11}},\ \bibinfo {pages}
  {10079} (\bibinfo {year} {1999})}\BibitemShut {NoStop}%
\bibitem [{\citenamefont {Lo~Verso}\ \emph {et~al.}(2006)\citenamefont
  {Lo~Verso}, \citenamefont {Vink}, \citenamefont {Pini},\ and\ \citenamefont
  {Reatto}}]{Loverso2006}%
  \BibitemOpen
  \bibfield  {author} {\bibinfo {author} {\bibfnamefont {F.}~\bibnamefont
  {Lo~Verso}}, \bibinfo {author} {\bibfnamefont {R.~L.~C.}\ \bibnamefont
  {Vink}}, \bibinfo {author} {\bibfnamefont {D.}~\bibnamefont {Pini}}, \ and\
  \bibinfo {author} {\bibfnamefont {L.}~\bibnamefont {Reatto}},\ }\bibfield
  {title} {\enquote {\bibinfo {title} {Critical behavior in colloid-polymer
  mixtures: Theory and simulation},}\ }\href {\doibase
  10.1103/PhysRevE.73.061407} {\bibfield  {journal} {\bibinfo  {journal} {Phys.
  Rev. E}\ }\textbf {\bibinfo {volume} {73}},\ \bibinfo {pages} {061407}
  (\bibinfo {year} {2006})}\BibitemShut {NoStop}%
\bibitem [{\citenamefont {Ashton}\ and\ \citenamefont
  {Wilding}(2014)}]{Ashton2014-jcp}%
  \BibitemOpen
  \bibfield  {author} {\bibinfo {author} {\bibfnamefont {D.~J.}\ \bibnamefont
  {Ashton}}\ and\ \bibinfo {author} {\bibfnamefont {N.~B.}\ \bibnamefont
  {Wilding}},\ }\bibfield  {title} {\enquote {\bibinfo {title} {Three-body
  interactions in complex fluids: Virial coefficients from simulation
  finite-size effects},}\ }\href
  {http://scitation.aip.org/content/aip/journal/jcp/140/24/10.1063/1.4883718}
  {\bibfield  {journal} {\bibinfo  {journal} {J. Chem. Phys.}\ }\textbf
  {\bibinfo {volume} {140}},\ \bibinfo {eid} {244118} (\bibinfo {year}
  {2014})}\BibitemShut {NoStop}%
\bibitem [{\citenamefont {Ashton}\ \emph {et~al.}(2011)\citenamefont {Ashton},
  \citenamefont {Wilding}, \citenamefont {Roth},\ and\ \citenamefont
  {Evans}}]{Ashton2011depletion}%
  \BibitemOpen
  \bibfield  {author} {\bibinfo {author} {\bibfnamefont {D.~J.}\ \bibnamefont
  {Ashton}}, \bibinfo {author} {\bibfnamefont {N.~B.}\ \bibnamefont {Wilding}},
  \bibinfo {author} {\bibfnamefont {R.}~\bibnamefont {Roth}}, \ and\ \bibinfo
  {author} {\bibfnamefont {R.}~\bibnamefont {Evans}},\ }\bibfield  {title}
  {\enquote {\bibinfo {title} {Depletion potentials in highly size-asymmetric
  binary hard-sphere mixtures: Comparison of simulation results with theory},}\
  }\href {\doibase 10.1103/PhysRevE.84.061136} {\bibfield  {journal} {\bibinfo
  {journal} {Phys. Rev. E}\ }\textbf {\bibinfo {volume} {84}},\ \bibinfo
  {pages} {061136} (\bibinfo {year} {2011})}\BibitemShut {NoStop}%
\bibitem [{\citenamefont {Dijkstra}\ and\ \citenamefont {van
  Roij}(2002)}]{Dijkstra2002}%
  \BibitemOpen
  \bibfield  {author} {\bibinfo {author} {\bibfnamefont {M.}~\bibnamefont
  {Dijkstra}}\ and\ \bibinfo {author} {\bibfnamefont {R.}~\bibnamefont {van
  Roij}},\ }\bibfield  {title} {\enquote {\bibinfo {title} {Entropic wetting
  and many-body induced layering in a model colloid-polymer mixture},}\ }\href
  {\doibase 10.1103/PhysRevLett.89.208303} {\bibfield  {journal} {\bibinfo
  {journal} {Phys. Rev. Lett.}\ }\textbf {\bibinfo {volume} {89}},\ \bibinfo
  {pages} {208303} (\bibinfo {year} {2002})}\BibitemShut {NoStop}%
\bibitem [{\citenamefont {Dijkstra}\ \emph {et~al.}(2006)\citenamefont
  {Dijkstra}, \citenamefont {van Roij}, \citenamefont {Roth},\ and\
  \citenamefont {Fortini}}]{Dijkstra2006}%
  \BibitemOpen
  \bibfield  {author} {\bibinfo {author} {\bibfnamefont {M.}~\bibnamefont
  {Dijkstra}}, \bibinfo {author} {\bibfnamefont {R.}~\bibnamefont {van Roij}},
  \bibinfo {author} {\bibfnamefont {R.}~\bibnamefont {Roth}}, \ and\ \bibinfo
  {author} {\bibfnamefont {A.}~\bibnamefont {Fortini}},\ }\bibfield  {title}
  {\enquote {\bibinfo {title} {Effect of many-body interactions on the bulk and
  interfacial phase behavior of a model colloid-polymer mixture},}\ }\href
  {https://link.aps.org/doi/10.1103/PhysRevE.73.041404} {\bibfield  {journal}
  {\bibinfo  {journal} {Phys. Rev. E}\ }\textbf {\bibinfo {volume} {73}},\
  \bibinfo {pages} {041404} (\bibinfo {year} {2006})}\BibitemShut {NoStop}%
\bibitem [{\citenamefont {Spyriouni}\ \emph {et~al.}(2007)\citenamefont
  {Spyriouni}, \citenamefont {Tzoumanekas}, \citenamefont {Theodorou},
  \citenamefont {Mueller-Plathe},\ and\ \citenamefont
  {Milano}}]{Spyriouni2007}%
  \BibitemOpen
  \bibfield  {author} {\bibinfo {author} {\bibfnamefont {T.}~\bibnamefont
  {Spyriouni}}, \bibinfo {author} {\bibfnamefont {C.}~\bibnamefont
  {Tzoumanekas}}, \bibinfo {author} {\bibfnamefont {D.}~\bibnamefont
  {Theodorou}}, \bibinfo {author} {\bibfnamefont {F.}~\bibnamefont
  {Mueller-Plathe}}, \ and\ \bibinfo {author} {\bibfnamefont {G.}~\bibnamefont
  {Milano}},\ }\bibfield  {title} {\enquote {\bibinfo {title} {Coarse-grained
  and reverse-mapped united-atom simulations of long-chain atactic polystyrene
  melts: structure, thermodynamic properties, chain conformations, and
  entanglements},}\ }\href {\doibase 10.1021/ma0700983} {\bibfield  {journal}
  {\bibinfo  {journal} {Macromolecules}\ }\textbf {\bibinfo {volume} {40}},\
  \bibinfo {pages} {3876} (\bibinfo {year} {2007})}\BibitemShut {NoStop}%
\bibitem [{\citenamefont {Vrij}(1976)}]{Vrij1976}%
  \BibitemOpen
  \bibfield  {author} {\bibinfo {author} {\bibfnamefont {A.}~\bibnamefont
  {Vrij}},\ }\bibfield  {title} {\enquote {\bibinfo {title} {Polymers at
  interfaces and the interactions in colloidal dispersions},}\ }\href@noop {}
  {\bibfield  {journal} {\bibinfo  {journal} {Pure Appl. Chem.}\ }\textbf
  {\bibinfo {volume} {48}},\ \bibinfo {pages} {471} (\bibinfo {year}
  {1976})}\BibitemShut {NoStop}%
\bibitem [{\citenamefont {Gast}, \citenamefont {Hall},\ and\ \citenamefont
  {Russel}(1983)}]{Gast1983}%
  \BibitemOpen
  \bibfield  {author} {\bibinfo {author} {\bibfnamefont {A.~P.}\ \bibnamefont
  {Gast}}, \bibinfo {author} {\bibfnamefont {C.~K.}\ \bibnamefont {Hall}}, \
  and\ \bibinfo {author} {\bibfnamefont {W.~B.}\ \bibnamefont {Russel}},\
  }\bibfield  {title} {\enquote {\bibinfo {title} {Polymer-induced phase
  separations in nonaqueous colloidal suspensions},}\ }\href
  {http://www.sciencedirect.com/science/article/pii/0021979783900279}
  {\bibfield  {journal} {\bibinfo  {journal} {J. Colloid. Interf. Sci.}\
  }\textbf {\bibinfo {volume} {96}},\ \bibinfo {pages} {251} (\bibinfo {year}
  {1983})}\BibitemShut {NoStop}%
\bibitem [{\citenamefont {Lekkerkerker}\ \emph {et~al.}(1992)\citenamefont
  {Lekkerkerker}, \citenamefont {Poon}, \citenamefont {Pusey}, \citenamefont
  {Stroobants},\ and\ \citenamefont {Warren}}]{Lekkerkerker1992}%
  \BibitemOpen
  \bibfield  {author} {\bibinfo {author} {\bibfnamefont {H.~N.~W.}\
  \bibnamefont {Lekkerkerker}}, \bibinfo {author} {\bibfnamefont {W.~C.-K.}\
  \bibnamefont {Poon}}, \bibinfo {author} {\bibfnamefont {P.~N.}\ \bibnamefont
  {Pusey}}, \bibinfo {author} {\bibfnamefont {A.}~\bibnamefont {Stroobants}}, \
  and\ \bibinfo {author} {\bibfnamefont {P.~B.}\ \bibnamefont {Warren}},\
  }\bibfield  {title} {\enquote {\bibinfo {title} {Phase behaviour of colloid
  {$+$} polymer mixtures},}\ }\href
  {https://doi.org/10.1209%2F0295-5075%2F20%2F6%2F015} {\bibfield  {journal}
  {\bibinfo  {journal} {Europhysics Letters ({EPL})}\ }\textbf {\bibinfo
  {volume} {20}},\ \bibinfo {pages} {559} (\bibinfo {year} {1992})}\BibitemShut
  {NoStop}%
\bibitem [{\citenamefont {Schmidt}\ \emph {et~al.}(2002)\citenamefont
  {Schmidt}, \citenamefont {{L\"{o}wen}}, \citenamefont {Brader},\ and\
  \citenamefont {Evans}}]{Schmidt2002}%
  \BibitemOpen
  \bibfield  {author} {\bibinfo {author} {\bibfnamefont {M.}~\bibnamefont
  {Schmidt}}, \bibinfo {author} {\bibfnamefont {H.}~\bibnamefont
  {{L\"{o}wen}}}, \bibinfo {author} {\bibfnamefont {J.~M.}\ \bibnamefont
  {Brader}}, \ and\ \bibinfo {author} {\bibfnamefont {R.}~\bibnamefont
  {Evans}},\ }\bibfield  {title} {\enquote {\bibinfo {title} {Density
  functional theory for a model colloid~polymer mixture: bulk fluid phases},}\
  }\href {https://doi.org/10.1088%2F0953-8984%2F14%2F40%2F323} {\bibfield
  {journal} {\bibinfo  {journal} {J. Phys.: Cond. Matt.}\ }\textbf {\bibinfo
  {volume} {14}},\ \bibinfo {pages} {9353} (\bibinfo {year}
  {2002})}\BibitemShut {NoStop}%
\bibitem [{\citenamefont {Brader}, \citenamefont {Evans},\ and\ \citenamefont
  {Schmidt}(2003)}]{Brader2003}%
  \BibitemOpen
  \bibfield  {author} {\bibinfo {author} {\bibfnamefont {J.~M.}\ \bibnamefont
  {Brader}}, \bibinfo {author} {\bibfnamefont {R.}~\bibnamefont {Evans}}, \
  and\ \bibinfo {author} {\bibfnamefont {M.}~\bibnamefont {Schmidt}},\
  }\bibfield  {title} {\enquote {\bibinfo {title} {Statistical mechanics of
  inhomogeneous model colloid---polymer mixtures},}\ }\href
  {https://doi.org/10.1080/0026897032000174263} {\bibfield  {journal} {\bibinfo
   {journal} {Mol. Phys.}\ }\textbf {\bibinfo {volume} {101}},\ \bibinfo
  {pages} {3349} (\bibinfo {year} {2003})}\BibitemShut {NoStop}%
\bibitem [{\citenamefont {Oversteegen}\ and\ \citenamefont
  {Roth}(2005)}]{Oversteegen2005}%
  \BibitemOpen
  \bibfield  {author} {\bibinfo {author} {\bibfnamefont {S.~M.}\ \bibnamefont
  {Oversteegen}}\ and\ \bibinfo {author} {\bibfnamefont {R.}~\bibnamefont
  {Roth}},\ }\bibfield  {title} {\enquote {\bibinfo {title} {General methods
  for free-volume theory},}\ }\href {\doibase 10.1063/1.1908765} {\bibfield
  {journal} {\bibinfo  {journal} {J. Chem. Phys.}\ }\textbf {\bibinfo {volume}
  {122}},\ \bibinfo {pages} {214502} (\bibinfo {year} {2005})}\BibitemShut
  {NoStop}%
\bibitem [{\citenamefont {Wilding}(1995)}]{wilding1995critical}%
  \BibitemOpen
  \bibfield  {author} {\bibinfo {author} {\bibfnamefont {N.~B.}\ \bibnamefont
  {Wilding}},\ }\bibfield  {title} {\enquote {\bibinfo {title} {Critical-point
  and coexistence-curve properties of the {L}ennard-{J}ones fluid: A
  finite-size scaling study},}\ }\href@noop {} {\bibfield  {journal} {\bibinfo
  {journal} {Phys. Rev. E}\ }\textbf {\bibinfo {volume} {52}},\ \bibinfo
  {pages} {602} (\bibinfo {year} {1995})}\BibitemShut {NoStop}%
\bibitem [{\citenamefont {Vink}\ and\ \citenamefont
  {Horbach}(2004)}]{Vink2004}%
  \BibitemOpen
  \bibfield  {author} {\bibinfo {author} {\bibfnamefont {R.~L.~C.}\
  \bibnamefont {Vink}}\ and\ \bibinfo {author} {\bibfnamefont {J.}~\bibnamefont
  {Horbach}},\ }\bibfield  {title} {\enquote {\bibinfo {title} {Grand canonical
  {M}onte {C}arlo simulation of a model colloid--polymer mixture: Coexistence
  line, critical behavior, and interfacial tension},}\ }\href {\doibase
  10.1063/1.1773771} {\bibfield  {journal} {\bibinfo  {journal} {J. Chem.
  Phys.}\ }\textbf {\bibinfo {volume} {121}},\ \bibinfo {pages} {3253}
  (\bibinfo {year} {2004})}\BibitemShut {NoStop}%
\bibitem [{\citenamefont {Oberhofer}\ and\ \citenamefont
  {Dellago}(2009)}]{Oberhofer2009}%
  \BibitemOpen
  \bibfield  {author} {\bibinfo {author} {\bibfnamefont {H.}~\bibnamefont
  {Oberhofer}}\ and\ \bibinfo {author} {\bibfnamefont {C.}~\bibnamefont
  {Dellago}},\ }\bibfield  {title} {\enquote {\bibinfo {title} {Efficient
  extraction of free energy profiles from nonequilibrium experiments},}\ }\href
  {\doibase 10.1002/jcc.21290} {\bibfield  {journal} {\bibinfo  {journal} {J.
  Comp. Chem.}\ }\textbf {\bibinfo {volume} {30}},\ \bibinfo {pages} {1726}
  (\bibinfo {year} {2009})}\BibitemShut {NoStop}%
\bibitem [{\citenamefont {Chen}(2005)}]{Chen2005}%
  \BibitemOpen
  \bibfield  {author} {\bibinfo {author} {\bibfnamefont {Y.}~\bibnamefont
  {Chen}},\ }\bibfield  {title} {\enquote {\bibinfo {title} {Another look at
  rejection sampling through importance sampling},}\ }\href
  {https://doi.org/10.1016/j.spl.2005.01.002} {\bibfield  {journal} {\bibinfo
  {journal} {Statist. Prob. Lett.}\ }\textbf {\bibinfo {volume} {72}},\
  \bibinfo {pages} {277} (\bibinfo {year} {2005})}\BibitemShut {NoStop}%
\bibitem [{\citenamefont {Agapiou}\ \emph {et~al.}(2017)\citenamefont
  {Agapiou}, \citenamefont {Papaspiliopoulos}, \citenamefont {Sanz-Alonso},\
  and\ \citenamefont {Stuart}}]{Agapiou2017}%
  \BibitemOpen
  \bibfield  {author} {\bibinfo {author} {\bibfnamefont {S.}~\bibnamefont
  {Agapiou}}, \bibinfo {author} {\bibfnamefont {O.}~\bibnamefont
  {Papaspiliopoulos}}, \bibinfo {author} {\bibfnamefont {D.}~\bibnamefont
  {Sanz-Alonso}}, \ and\ \bibinfo {author} {\bibfnamefont {A.~M.}\ \bibnamefont
  {Stuart}},\ }\bibfield  {title} {\enquote {\bibinfo {title} {Importance
  sampling: Intrinsic dimension and computational cost},}\ }\href {\doibase
  10.1214/17-STS611} {\bibfield  {journal} {\bibinfo  {journal} {Statist.
  Sci.}\ }\textbf {\bibinfo {volume} {32}},\ \bibinfo {pages} {405} (\bibinfo
  {year} {2017})}\BibitemShut {NoStop}%
\bibitem [{\citenamefont {Liu}, \citenamefont {Wilding},\ and\ \citenamefont
  {Luijten}(2006)}]{Liu2006}%
  \BibitemOpen
  \bibfield  {author} {\bibinfo {author} {\bibfnamefont {J.}~\bibnamefont
  {Liu}}, \bibinfo {author} {\bibfnamefont {N.~B.}\ \bibnamefont {Wilding}}, \
  and\ \bibinfo {author} {\bibfnamefont {E.}~\bibnamefont {Luijten}},\
  }\bibfield  {title} {\enquote {\bibinfo {title} {Simulation of phase
  transitions in highly asymmetric fluid mixtures},}\ }\href@noop {} {\bibfield
   {journal} {\bibinfo  {journal} {Phys. Rev. Lett.}\ }\textbf {\bibinfo
  {volume} {97}} (\bibinfo {year} {2006})}\BibitemShut {NoStop}%
\bibitem [{\citenamefont {Ashton}\ and\ \citenamefont
  {Wilding}(2011)}]{Ashton2011}%
  \BibitemOpen
  \bibfield  {author} {\bibinfo {author} {\bibfnamefont {D.~J.}\ \bibnamefont
  {Ashton}}\ and\ \bibinfo {author} {\bibfnamefont {N.~B.}\ \bibnamefont
  {Wilding}},\ }\bibfield  {title} {\enquote {\bibinfo {title} {Grand canonical
  simulation of phase behaviour in highly size-asymmetrical binary fluids},}\
  }\href {\doibase 10.1080/00268976.2010.482067} {\bibfield  {journal}
  {\bibinfo  {journal} {Mol. Phys.}\ }\textbf {\bibinfo {volume} {109}},\
  \bibinfo {pages} {999} (\bibinfo {year} {2011})}\BibitemShut {NoStop}%
\bibitem [{\citenamefont {Dijkstra}, \citenamefont {van Roij},\ and\
  \citenamefont {Evans}(1998)}]{Dijkstra1998}%
  \BibitemOpen
  \bibfield  {author} {\bibinfo {author} {\bibfnamefont {M.}~\bibnamefont
  {Dijkstra}}, \bibinfo {author} {\bibfnamefont {R.}~\bibnamefont {van Roij}},
  \ and\ \bibinfo {author} {\bibfnamefont {R.}~\bibnamefont {Evans}},\
  }\bibfield  {title} {\enquote {\bibinfo {title} {Phase behavior and structure
  of binary hard-sphere mixtures},}\ }\href {\doibase
  10.1103/PhysRevLett.81.2268} {\bibfield  {journal} {\bibinfo  {journal}
  {Phys. Rev. Lett.}\ }\textbf {\bibinfo {volume} {81}},\ \bibinfo {pages}
  {2268} (\bibinfo {year} {1998})}\BibitemShut {NoStop}%
\bibitem [{\citenamefont {Dijkstra}, \citenamefont {van Roij},\ and\
  \citenamefont {Evans}(1999)}]{Dijkstra1999-pre}%
  \BibitemOpen
  \bibfield  {author} {\bibinfo {author} {\bibfnamefont {M.}~\bibnamefont
  {Dijkstra}}, \bibinfo {author} {\bibfnamefont {R.}~\bibnamefont {van Roij}},
  \ and\ \bibinfo {author} {\bibfnamefont {R.}~\bibnamefont {Evans}},\
  }\bibfield  {title} {\enquote {\bibinfo {title} {Phase diagram of highly
  asymmetric binary hard-sphere mixtures},}\ }\href {\doibase
  10.1103/PhysRevE.59.5744} {\bibfield  {journal} {\bibinfo  {journal} {Phys.
  Rev. E}\ }\textbf {\bibinfo {volume} {59}},\ \bibinfo {pages} {5744}
  (\bibinfo {year} {1999})}\BibitemShut {NoStop}%
\bibitem [{\citenamefont {Odriozola}, \citenamefont {Jimenez-Angeles},\ and\
  \citenamefont {Lozada-Cassou}(2008)}]{Odriozola2008}%
  \BibitemOpen
  \bibfield  {author} {\bibinfo {author} {\bibfnamefont {G.}~\bibnamefont
  {Odriozola}}, \bibinfo {author} {\bibfnamefont {F.}~\bibnamefont
  {Jimenez-Angeles}}, \ and\ \bibinfo {author} {\bibfnamefont {M.}~\bibnamefont
  {Lozada-Cassou}},\ }\bibfield  {title} {\enquote {\bibinfo {title} {Entropy
  driven key-lock assembly},}\ }\href {\doibase DOI 10.1063/1.2981795}
  {\bibfield  {journal} {\bibinfo  {journal} {J. Chem. Phys.}\ }\textbf
  {\bibinfo {volume} {129}},\ \bibinfo {pages} {111101} (\bibinfo {year}
  {2008})}\BibitemShut {NoStop}%
\bibitem [{\citenamefont {van Anders}\ \emph {et~al.}(2014)\citenamefont {van
  Anders}, \citenamefont {Klotsa}, \citenamefont {Ahmed}, \citenamefont
  {Engel},\ and\ \citenamefont {Glotzer}}]{vanAnders2014}%
  \BibitemOpen
  \bibfield  {author} {\bibinfo {author} {\bibfnamefont {G.}~\bibnamefont {van
  Anders}}, \bibinfo {author} {\bibfnamefont {D.}~\bibnamefont {Klotsa}},
  \bibinfo {author} {\bibfnamefont {N.~K.}\ \bibnamefont {Ahmed}}, \bibinfo
  {author} {\bibfnamefont {M.}~\bibnamefont {Engel}}, \ and\ \bibinfo {author}
  {\bibfnamefont {S.~C.}\ \bibnamefont {Glotzer}},\ }\bibfield  {title}
  {\enquote {\bibinfo {title} {Understanding shape entropy through local dense
  packing},}\ }\href {\doibase 10.1073/pnas.1418159111} {\bibfield  {journal}
  {\bibinfo  {journal} {Proc. Natl. Acad. Sci. USA}\ }\textbf {\bibinfo
  {volume} {111}},\ \bibinfo {pages} {E4812} (\bibinfo {year}
  {2014})}\BibitemShut {NoStop}%
\bibitem [{\citenamefont {Law}\ \emph {et~al.}(2016)\citenamefont {Law},
  \citenamefont {Ashton}, \citenamefont {Wilding},\ and\ \citenamefont
  {Jack}}]{Law2016}%
  \BibitemOpen
  \bibfield  {author} {\bibinfo {author} {\bibfnamefont {C.}~\bibnamefont
  {Law}}, \bibinfo {author} {\bibfnamefont {D.~J.}\ \bibnamefont {Ashton}},
  \bibinfo {author} {\bibfnamefont {N.~B.}\ \bibnamefont {Wilding}}, \ and\
  \bibinfo {author} {\bibfnamefont {R.~L.}\ \bibnamefont {Jack}},\ }\bibfield
  {title} {\enquote {\bibinfo {title} {Coarse-grained depletion potentials for
  anisotropic colloids: Application to lock-and-key systems},}\ }\href
  {\doibase 10.1063/1.4961541} {\bibfield  {journal} {\bibinfo  {journal} {J.
  Chem. Phys.}\ }\textbf {\bibinfo {volume} {145}},\ \bibinfo {pages} {084907}
  (\bibinfo {year} {2016})}\BibitemShut {NoStop}%
\bibitem [{\citenamefont {Ashton}, \citenamefont {Jack},\ and\ \citenamefont
  {Wilding}(2015)}]{Ashton2015}%
  \BibitemOpen
  \bibfield  {author} {\bibinfo {author} {\bibfnamefont {D.~J.}\ \bibnamefont
  {Ashton}}, \bibinfo {author} {\bibfnamefont {R.~L.}\ \bibnamefont {Jack}}, \
  and\ \bibinfo {author} {\bibfnamefont {N.~B.}\ \bibnamefont {Wilding}},\
  }\bibfield  {title} {\enquote {\bibinfo {title} {Porous liquid phases for
  indented colloids with depletion interactions},}\ }\href {\doibase
  10.1103/PhysRevLett.114.237801} {\bibfield  {journal} {\bibinfo  {journal}
  {Phys. Rev. Lett.}\ }\textbf {\bibinfo {volume} {114}},\ \bibinfo {pages}
  {237801} (\bibinfo {year} {2015})}\BibitemShut {NoStop}%
\bibitem [{\citenamefont {Jusufi}, \citenamefont {Watzlawek},\ and\
  \citenamefont {L{\"o}wen}(1999)}]{Jusufi1999}%
  \BibitemOpen
  \bibfield  {author} {\bibinfo {author} {\bibfnamefont {A.}~\bibnamefont
  {Jusufi}}, \bibinfo {author} {\bibfnamefont {M.}~\bibnamefont {Watzlawek}}, \
  and\ \bibinfo {author} {\bibfnamefont {H.}~\bibnamefont {L{\"o}wen}},\
  }\bibfield  {title} {\enquote {\bibinfo {title} {Effective interaction
  between star polymers},}\ }\href {https://doi.org/10.1021/ma981844u}
  {\bibfield  {journal} {\bibinfo  {journal} {Macromolecules}\ }\textbf
  {\bibinfo {volume} {32}},\ \bibinfo {pages} {4470} (\bibinfo {year}
  {1999})}\BibitemShut {NoStop}%
\bibitem [{\citenamefont {Marzi}, \citenamefont {Likos},\ and\ \citenamefont
  {Capone}(2012)}]{Marzi2012}%
  \BibitemOpen
  \bibfield  {author} {\bibinfo {author} {\bibfnamefont {D.}~\bibnamefont
  {Marzi}}, \bibinfo {author} {\bibfnamefont {C.~N.}\ \bibnamefont {Likos}}, \
  and\ \bibinfo {author} {\bibfnamefont {B.}~\bibnamefont {Capone}},\
  }\bibfield  {title} {\enquote {\bibinfo {title} {Coarse graining of
  star-polymer--colloid nanocomposites},}\ }\href {\doibase 10.1063/1.4730751}
  {\bibfield  {journal} {\bibinfo  {journal} {J. Chem. Phys.}\ }\textbf
  {\bibinfo {volume} {137}},\ \bibinfo {pages} {014902} (\bibinfo {year}
  {2012})}\BibitemShut {NoStop}%
\bibitem [{\citenamefont {Lenz}\ \emph {et~al.}(2012)\citenamefont {Lenz},
  \citenamefont {Blaak}, \citenamefont {Likos},\ and\ \citenamefont
  {Mladek}}]{Lenz2012}%
  \BibitemOpen
  \bibfield  {author} {\bibinfo {author} {\bibfnamefont {D.~A.}\ \bibnamefont
  {Lenz}}, \bibinfo {author} {\bibfnamefont {R.}~\bibnamefont {Blaak}},
  \bibinfo {author} {\bibfnamefont {C.~N.}\ \bibnamefont {Likos}}, \ and\
  \bibinfo {author} {\bibfnamefont {B.~M.}\ \bibnamefont {Mladek}},\ }\bibfield
   {title} {\enquote {\bibinfo {title} {Microscopically resolved simulations
  prove the existence of soft cluster crystals},}\ }\href {\doibase
  10.1103/PhysRevLett.109.228301} {\bibfield  {journal} {\bibinfo  {journal}
  {Phys. Rev. Lett.}\ }\textbf {\bibinfo {volume} {109}},\ \bibinfo {pages}
  {228301} (\bibinfo {year} {2012})}\BibitemShut {NoStop}%
\bibitem [{\citenamefont {Wilding}\ and\ \citenamefont
  {Sollich}(2014)}]{Wilding2014}%
  \BibitemOpen
  \bibfield  {author} {\bibinfo {author} {\bibfnamefont {N.~B.}\ \bibnamefont
  {Wilding}}\ and\ \bibinfo {author} {\bibfnamefont {P.}~\bibnamefont
  {Sollich}},\ }\bibfield  {title} {\enquote {\bibinfo {title} {Demixing
  cascades in cluster crystals},}\ }\href {\doibase 10.1063/1.4894374}
  {\bibfield  {journal} {\bibinfo  {journal} {J. Chem. Phys.}\ }\textbf
  {\bibinfo {volume} {141}},\ \bibinfo {pages} {094903} (\bibinfo {year}
  {2014})}\BibitemShut {NoStop}%
\bibitem [{\citenamefont {Largo}\ and\ \citenamefont
  {Wilding}(2006)}]{Largo2006}%
  \BibitemOpen
  \bibfield  {author} {\bibinfo {author} {\bibfnamefont {J.}~\bibnamefont
  {Largo}}\ and\ \bibinfo {author} {\bibfnamefont {N.~B.}\ \bibnamefont
  {Wilding}},\ }\bibfield  {title} {\enquote {\bibinfo {title} {Influence of
  polydispersity on the critical parameters of an effective-potential model for
  asymmetric hard-sphere mixtures},}\ }\href {\doibase
  10.1103/PhysRevE.73.036115} {\bibfield  {journal} {\bibinfo  {journal} {Phys.
  Rev. E}\ }\textbf {\bibinfo {volume} {73}},\ \bibinfo {pages} {036115}
  (\bibinfo {year} {2006})}\BibitemShut {NoStop}%
\end{thebibliography}%

\end{document}